\documentclass[12pt]{article}
\usepackage[dvipdfmx]{graphicx,color}
\usepackage{amsmath,amssymb,accents}
\usepackage{amsfonts}
\usepackage{graphicx,color}
\usepackage{cite}
\usepackage{bm}
\usepackage{comment}

\newcommand{\comma}{ \, , }
\newcommand{\period}{ \, . }
\newcommand{\del}{  \partial  }
\newcommand{\delbar}{  \bar{\partial}  }
\newcommand{\zbar}{\bar{z}}

\newcommand{\nn}{\nonumber}
\newcommand{\be}         {  \begin{equation}  }
\newcommand{\ee}           {  \end{equation}  }
\newcommand{\eqn}[1]{(\ref{#1})}

\newcommand{\bbR}{{\mathbb R}}
\newcommand{\bbZ}{{\mathbb Z}}

\renewcommand{\thefootnote}{\fnsymbol{footnote}}

\newcommand{\Exp}[1]{\operatorname{e}^{#1}}
\newcommand{\rmd}{{\mathrm{d}}}

\newcommand{\cB}{\mathcal B}
\newcommand{\cD}{\mathcal D}
\newcommand{\cE}{\mathcal E}
\newcommand{\cH}{\mathcal H}
\newcommand{\cJ}{\mathcal J}

\newcommand{\cR}{\mathcal R}

\newcommand{\cZ}{\mathcal Z}

\newcommand{\OO}{\text{O}}

\newcommand{\hh}{h}

\newcommand{\gga}{\alpha}
\newcommand{\ggb}{\beta}
\newcommand{\ggc}{\gamma}
\newcommand{\ggd}{\delta}

\makeatletter
\newcommand{\ubar}[1]{\underaccent{\bar}{#1}}
\makeatother

\makeatletter \@addtoreset{equation}{section} \makeatother
\renewcommand{\theequation}{\arabic{section}.\arabic{equation}}

\renewcommand{\thefootnote}{\arabic{footnote}}
\begin{document}
\thispagestyle{empty}
\setcounter{page}{0}
\baselineskip=3.5ex
\vspace*{4.5ex}
\begin{center}
\Large\textbf{On quantum Poisson--Lie T-duality of WZNW models}
\end{center}

\vspace{7mm}

\centerline{
{Yuho Sakatani}%
\footnote{E-mail address: yuho@koto.kpu-m.ac.jp}
and
{Yuji Satoh}%
\footnote{E-mail address: ysatoh@u-fukui.ac.jp}
}

\vspace*{1ex}
\begin{center}
${}^\ast {}^\ddagger${\it Department of Physics, Kyoto Prefectural University of Medicine,}\\
{\it 1-5 Shimogamohangi-cho, Sakyo-ku, Kyoto 606-0823, Japan}

\vspace*{1ex}
${}^\dagger${\it Department of Applied Physics, University of Fukui,}\\
{\it Bunkyo 3-9-1, Fukui 910-8507, Japan\\}
\end{center}

\vspace*{4ex}
\begin{abstract}
\baselineskip=3.4ex
We study Poisson--Lie T-duality of the Wess--Zumino--Novikov--Witten (WZNW) models which are obtained from a class of Drinfel'd doubles and its generalization. In this case, the resultant WZNW models are known to be classically self-dual under Poisson--Lie T-duality. We describe an explicit construction of the associated currents, and discuss the conformal invariance under this duality. In a concrete example of the SU(2) WZNW model, we find that the self-duality is represented as a chiral automorphism of the $\widehat{\mathfrak{su}}(2)$ affine Lie algebra, though the transformation of the currents is non-local and non-linear. This classical automorphism can be promoted to the quantum one through the parafermionic formulation of $\widehat{\mathfrak{su}}(2)$, which in turn induces an isomorphism of the WZNW model. We thus find a full quantum equivalence of the dual pair under Poisson--Lie T-duality. The isomorphism is represented by a sign-change of a chiral boson or the order-disorder duality of the parafermionic conformal field theory as in Abelian T-duality on tori or in the mirror symmetry of the Gepner model. 
\end{abstract}

\setlength{\parskip}{1ex}
\renewcommand{\thefootnote}{\arabic{footnote}}
\setcounter{footnote}{0}
\setcounter{section}{0}
\pagestyle{plain}

\baselineskip=3.6ex
\newpage

\section{Introduction}
\label{sec:introduction}

T-duality of string theory \cite{Kikkawa:1984cp,Sakai:1985cs} is a duality between string theories defined on different target spaces.
This is known as the exact/quantum duality of string theory on tori $T^d$, where the duality group forms O$(d,d;\bbZ)$\,. 
The duality also gives rise to an O$(d,d;\bbR)$ symmetry of the supergravity equations of motion that describe the strings on $T^d$ at low energy \cite{Buscher:1987sk}. 
This O$(d,d)$ symmetry has been studied from various perspectives in string sigma models 
(see for example \cite{Duff:1989tf,Giveon:1991jj,Rocek:1991ps,Alvarez:1993qi,Alvarez:1994wj}). 
In particular, a derivation of the duality as an isomorphism of conformal field theories (CFTs) has been given in \cite{Rocek:1991ps}. 
The global structure of the dual geometry in general cases has been discussed in detail in \cite{Alvarez:1993qi}. 

This well-established T-duality requires the existence of $d$ commuting (Abelian) isometries and is called Abelian T-duality. 
The existence of Abelian isometries gives a strong restriction on the target space, and extensions of T-duality that does not require Abelian isometries have been proposed. 
The first of such extensions has been known as non-Abelian T-duality \cite{delaOssa:1992vci,Giveon:1993ai,Alvarez:1994np}, which assumes the existence of non-Abelian isometries in the target space. 
Because the dual geometry does not always have isometries, this map is generally a one-way map, and we cannot recover the original model from the dual one. 
This situation has been improved in \cite{Klimcik:1995ux,Klimcik:1995jn} and the generalized duality is called Poisson--Lie T-duality. 
This duality assumes the existence of certain geometrical structure, Poisson--Lie symmetry, in the target space, and the standard isometries are not necessary. 
Accordingly, the resultant non-Abelian duality has become a true ``duality" in both directions, and more general extensions of T-duality have become possible.
Similarly to Abelian T-duality in general, Poisson--Lie T-duality is a classical symmetry of string sigma models \cite{Klimcik:1995dy,Sfetsos:1997pi}, and it is also a symmetry of the supergravity equations of motion (see for example \cite{Bossard:2001au}). 
Thus Poisson--Lie T-duality is frequently employed as a solution-generating technique in supergravity. 

To study the symmetry of the supergravity equations of motion, it is useful to employ T-dual-manifest formulations of supergravity, such as double field theory (DFT) \cite{Siegel:1993th,Hull:2009mi,Hohm:2010pp,Geissbuhler:2013uka}. 
Using these frameworks, we can clearly show the duality covariance of the supergravity equations of motion even in the presence of the Ramond--Ramond fields or the spectator fields (or the external spacetime that is invariant under the duality) \cite{Hassler:2017yza,Severa:2018pag,Demulder:2018lmj,Sakatani:2019jgu}. 
Poisson--Lie T-duality is based on a $2d$-dimensional Lie algebra, and if the structure constants of its $d$-dimensional subalgebra satisfy $f_a{}^{ab}\neq 0$, it is known that the dual geometry, in general, satisfies the generalized supergravity equations of motion \cite{Arutyunov:2015mqj}, which are shown in \cite{Wulff:2016tju} to be equivalent to the $\kappa$-symmetry of the Green--Schwarz superstring.

In that case, the conformal symmetry may be broken in the dual model, as is found in \cite{Gasperini:1993nz}.
However, the generalized supergravity equations of motion can be derived from the DFT equations of motion by choosing a non-standard section, and then, as has been discussed in \cite{Sakamoto:2017wor}, we can always consider the corresponding background as a solution of DFT. 
By performing a suitable redefinition of the coordinates and fields that maps the configuration into the standard section, we get another solution which solves the standard supergravity equations of motion where the conformal invariance is recovered even when $f_a{}^{ab}\neq 0$. 
Moreover, in some particular cases, it is possible that the dual geometry satisfies the standard supergravity equations of motion without any redefinition (or with a shift of the dilaton) \cite{Wulff:2018aku}. 
Poisson--Lie T-duality can be understood as a duality map between conformal string sigma models also in this sense.

Unlike well-established Abelian T-duality, several open questions remain in the case of Poisson--Lie T-duality.
One is the global structure of the dual model. 
For example, there is no established way as in \cite{Alvarez:1993qi} to discuss the global structure, such as the period, of the dual geometry. 
Another is its quantum aspects.
In this respect, loop corrections of Poisson--Lie symmetric models have been studied for example in \cite{Balog:1998br,Sfetsos:1999zm,Sfetsos:2009dj,Sfetsos:2009vt}. 
Also from the point of view of DFT, Poisson--Lie T-duality in the presence of the $\alpha'$-corrections has been studied in \cite{Hassler:2020tvz,Borsato:2020wwk,Codina:2020yma}.
Path integral formulations of Poisson--Lie T-duality have been discussed in \cite{Alekseev:1995ym,Tyurin:1995bu,VonUnge:2002xjf}. 
However, the quantum equivalence of the dual pairs under Poisson--Lie T-duality is generally yet to be understood.

To take a step toward understanding these issues, in this paper we conduct a detailed study of Poisson--Lie T-duality of Wess--Zumino--Novikov--Witten (WZNW) models. 
Under certain mild conditions, which are discussed in section \ref{sec:SD-WZNW}, WZNW models are known to be self-dual under Poisson--Lie T-duality \cite{Klimcik:1996hp}, and we focus on this case. 
We describe an explicit construction of the associated currents.
The conformal invariance in this self-duality is also discussed.

As concrete examples, we consider a certain six-dimensional Drinfel'd double and its generalization in section \ref{sec:PLWZW}, which result in the self-duality of the SU(2) WZNW model. 
We find an explicit relation between the dual pairs of the currents.
Though this relation is non-local, non-linear and involving an infinite number of the modes of the currents, it turns out to be a chiral automorphism of the associated affine Lie algebra $\widehat{\mathfrak{su}}(2)$ in terms of the Poisson bracket.
This is also understood as a consequence of the general property that Poisson--Lie T-duality is represented by a canonical transformation \cite{Klimcik:1995dy,Sfetsos:1997pi}.

We then discuss the quantum aspects, as well as the global structure, of this self-duality in section \ref{sec:quantum}.
A quantum equivalence under any duality may be reduced to an isomorphism of the underlying CFT, as in the case of Abelian T-duality on tori \cite{Rocek:1991ps}.
We find that the above classical automorphism can be promoted to the quantum one through the parafermionic formulation of $\widehat{\mathfrak{su}}(2)$, which indeed induces an isomorphism of the WZNW model. 
This establishes the full quantum equivalence of the dual pair in our case.
The isomorphism of the CFT implies that the equivalence holds to all orders in $\alpha'$ and at any genus of the world-sheet.
In the course, the global structure of the duality is also figured out. 
The isomorphism is represented by a change in the sign of the associated chiral free boson or the order-disorder duality of the parafermionic CFT \cite{Fateev:1985mm,Gepner:1986hr}, as in Abelian T-duality on tori or in the mirror symmetry \cite{Greene:1990ud,Greene:1996cy} of the Gepner model \cite{Gepner:1987qi}. 
Equivalently, the isomorphism is understood as that between the SU(2) WZNW model and its orbifold model
\cite{Gepner:1986hr,Gaberdiel:1995mx,Maldacena:2001ky}.

The organization of the rest of this paper is as follows.
In section \ref{sec:SD-WZNW}, we discuss general properties of the self-duality of the WZNW models under Poisson--Lie T-duality.
In section \ref{sec:PLWZW}, we consider concrete examples of the self-duality of the SU(2) WZNW model, 
and find explicit relations between the dual pairs. 
In section \ref{sec:quantum}, we discuss the quantum aspects, as well as the global structure, of the self-duality. 
We also observe that the self-duality in our case is represented by Abelian T-duality of WZNW models \cite{Kiritsis:1993ju,Alvarez:1993qi,Alvarez:1994wj}.
We conclude with a summary and discussion in section \ref{sec:conclusions}.

\section{Poisson--Lie T-duality and self-duality of WZNW models}
\label{sec:SD-WZNW}

Poisson--Lie T-duality \cite{Klimcik:1995ux,Klimcik:1995jn} is a (classical) equivalence of two or more string sigma models on some cosets $\cD/\tilde{G}$, $\cD/\tilde{G}'$, $\cdots$. 
When $\mathfrak{r}$ is a $d$-dimensional Lie subalgebra of $\mathfrak{d} = {\rm Lie} (\cD)$ transversal to $\tilde{\mathfrak{g}}={\rm Lie}(\tilde{G})$ and $\tilde{\mathfrak{g}}'={\rm Lie}(\tilde{G}')$ and $R = \exp(\mathfrak{r})$ is a compact Lie subgroup of $\cD$, then both cosets $\cD/\tilde{G}$ and $\cD/\tilde{G}'$ can be identified with $R$ or its discrete coset, and there is a choice of the dual pair of sigma models on $\cD/\tilde{G} $ and $\cD/\tilde{G}' $ 
such that both models are the WZNW models associated with $R$ \cite{Klimcik:1996hp}.
In this case, Poisson--Lie T-duality becomes self-dual at the level of equations of motion and we find a non-trivial map between the WZNW models with the same $R$.
Here we review the classical aspects of this self-duality, extending the Drinfel'd double, which corresponds to $\cD$, by a slightly more general algebra. 
We also study aspects of WZNW models using the formulation of DFT. 
The classical analyses in this and the next section apply up to global issues, which are discussed in detail in section \ref{sec:quantum}.
For Poisson-Lie T-duality in WZNW models, see for example \cite{Alekseev:1995ym,Eghbali:2015yoa,Eghbali:2018ohx,Sakatani:2021eqt}.

\subsection{Set-up}
We consider a $2d$-dimensional Lie algebra $\mathfrak{d}$\,,
\begin{align}
\label{eq:LieAlgD}
 [T_A,\,T_B] = F_{AB}{}^C\,T_C\,,
\end{align}
which admits an adjoint-invariant bilinear form of split signature $(d,d)$
\begin{align}
 \eta_{AB} \equiv \langle T_A,\,T_B\rangle\,. 
\end{align}
We assume that $\mathfrak{d}$ contains a maximally (i.e., $d$-dimensional) isotropic subalgebra $\tilde{\mathfrak{g}}$\,. 
We parameterize the generators as $T_A=(T_a,\,T^a)$ ($a=1,\dotsc,d$) and choose $T^a$ to be the generators of $\tilde{\mathfrak{g}}$\,. 
The other generators $T_a$ are chosen such that the bilinear form takes the form
\begin{align}
 \eta_{AB} &= \begin{pmatrix} 0 & \delta_a^b \\ \delta^a_b & 0 \end{pmatrix}.
\end{align}
We always raise or lower the indices $A$ by using $\eta_{AB}$\,, and then the adjoint invariance shows that $F_{ABC}\equiv F_{AB}{}^D\,\eta_{DC}$ is totally antisymmetric. 
Then we can express the Lie algebra $\mathfrak{d}$ as
\begin{align}
\begin{split}
 [T_a,\,T_b] &= f_{ab}{}^c\,T_c + f_{abc}\,T^c\,,\qquad
 [T^a,\,T^b] = f_c{}^{ab}\,T^c\,,
\\
 [T_a,\,T^b] &= f_a{}^{bc} \,T_c - f_{ac}{}^b\,T^c = - [T^b,\,T_a]\,,
 \label{eq:Tcommrel}
\end{split}
\end{align}
where $f_{ab}{}^c=-f_{ba}{}^c$, $f_a{}^{bc}=-f_a{}^{cb}$, and $f_{abc}=f_{[abc]}$, and $f_a{}^{bc}$ corresponds to the structure constants of $\tilde{\mathfrak{g}}$\,. 
If the structure constants $f_{abc}$ are absent this reduces to the Lie algebra of the Drinfel'd double, but their absence is not assumed in the following discussion. 

We now consider the motion of the strings on the Lie group $\cD=\exp(\mathfrak{d})$. 
The embedding of the string world-sheet $\Sigma$ is described by $l(\sigma)\in \cD$\,, and the action can be given by \cite{Hull:2009sg,Reid-Edwards:2010plo}
\begin{align}
 4\pi\tilde{\alpha}'\,S= \frac{1}{2}\int_\Sigma \langle \rmd l\,l^{-1} ,\, * \hat{\cH}(\rmd l\,l^{-1})\rangle -\frac{1}{3!}\int_{\cB} \langle \rmd l\,l^{-1},\,[\rmd l\,l^{-1},\,\rmd l\,l^{-1}]\rangle \,,
\label{eq:SM-action}
\end{align}
where $\partial \cB=\Sigma$ and $\hat{\cH}_{AB}\in \OO(d,d)$ is a constant symmetric matrix that may be parameterized as
\begin{align}
 \hat{\cH}_{AB} = \begin{pmatrix} [\hat{g} - \hat{B}\,(\hat{g})^{-1}\,\hat{B}]_{ab} & [\hat{B}\,(\hat{g})^{-1}]_a{}^b \\ -[(\hat{g})^{-1}\,\hat{B}]^a{}_b & [(\hat{g})^{-1}]^{ab} 
\label{eq:cH-param}\end{pmatrix}.
\end{align}
We have introduced a dimensionless parameter $\tilde{\alpha}'\equiv \alpha'/L^2$ where $L$ is a length scale of the target space. 
To reproduce the conventional string theory, we impose the self-duality constraint \cite{Hull:2004in,Hull:2009sg} at the level of equations of motion
\begin{align}
 \hat{\cH}(\rmd l\,l^{-1}) = * \rmd l\,l^{-1}\,.
\label{eq:SD}
\end{align}

\subsection{Conventional string sigma model}
Here we clarify the relation to the conventional string sigma model, by following section 3.3.1 of \cite{Reid-Edwards:2010plo} but without assuming $f_{abc}=0$\,. 
We choose a maximally isotropic subgroup $\tilde{G}=\exp(\tilde{\mathfrak{g}})$ of $\cD$ and decompose the group element $l(\sigma)\in\cD$ as
\begin{align}
 l(\sigma) = f(\sigma)\,\tilde{g}(\sigma) \,,
\end{align}
where $f(\sigma)\in \cD/\tilde{G}$ and $\tilde{g}\in \tilde{G}$ (recall that $T^a$ generate $\tilde{\mathfrak{g}}$ in our convention).
We shall define
\begin{align}
\begin{split}
 \ell &\equiv f^{-1}\,\rmd f \equiv \ell^A\,T_A \,,\qquad
 r \equiv \rmd f\,f^{-1}\equiv r^A\,T_A\,,
\\
 \tilde{\ell}&\equiv \tilde{g}^{-1}\,\rmd \tilde{g} \equiv \tilde{\ell}_a\,T^a \,,\qquad
 \tilde{r}\equiv \rmd \tilde{g}\,\tilde{g}^{-1}\equiv \tilde{r}_a\,T^a \,,
\end{split}
\end{align}
and then $\rmd l \,l^{-1}$ can be expressed as
\begin{align}
 \rmd l \,l^{-1} = r + f\,\tilde{r}\,f^{-1} = f\,\bigl(\ell + \tilde{r}\bigr)\,f^{-1} \,.
\end{align}
Using the self-duality constraint \eqref{eq:SD}, we can express $\tilde{r}$ by using $\ell$ as
\begin{align}
 \tilde{r}_a = (\rmd\tilde{g}\,\tilde{g}^{-1})_a = - \ell_a + B^f_{ab}\,\ell^b + g^f_{ab}*\ell^b \,.
 \label{eq:r-tilde}
\end{align}
Then eliminating $\tilde{r}$, the action \eqref{eq:SM-action} can be expressed as
\begin{align}
 S = \frac{1}{4\pi\tilde{\alpha}'}\int_{\Sigma} \bigl(g^f_{ab}\,\ell^a \wedge *\ell^b + B^f_{ab}\,\ell^a \wedge \ell^b - \ell^a \wedge \ell_a \bigr)
 -\frac{1}{24\pi\tilde{\alpha}'}\int_{\cB} \langle \ell ,\, [\ell ,\, \ell]\rangle \,.
\label{eq:SM-action22}
\end{align}
Here we have defined
\begin{align}
 f^{-1}\,T_A\,f \equiv M_A{}^B\,T_B\,,\qquad
 \cH^f_{AB}\equiv (M^{-1})_A{}^C\,(M^{-1})_B{}^D\,\hat{\cH}_{CD}\,,
\end{align}
and parameterized $\cH^f_{AB}$ as
\begin{align}
 \cH^f_{AB} = \begin{pmatrix} [g^f - B^f\,(g^{f})^{-1}\,B^f]_{ab} & [B^f\,(g^{f})^{-1}]_a{}^b \\ -[(g^f)^{-1}\,B^f]^a{}_b & [(g^f)^{-1}]^{ab} \end{pmatrix}.
\end{align}
The superscript $f$ is a mnemonic for the adjoint action by $f$.
If we further rewrite the action \eqref{eq:SM-action22} as
\begin{align}
 S = \frac{1}{4\pi\tilde{\alpha}'}\int_{\Sigma} \bigl(g_{mn}\,\rmd x^m \wedge *\rmd x^n + B_{mn}\,\rmd x^m \wedge \rmd x^n \bigr) \,,
\end{align}
the background fields can be determined as
\begin{align}
 g_{mn}\equiv g^f_{ab}\,\ell^a_m\,\ell^b_n\,,\quad
 B_{mn}\equiv B^f_{ab}\,\ell^a_m\,\ell^b_n - \ell^a_{[m|}\,\ell_{a|n]} + B^{\text{\tiny WZ}}_{mn}\,,\quad
 \rmd B^{\text{\tiny WZ}} \equiv - \tfrac{1}{2}\,\langle \ell ,\, [\ell ,\, \ell]\rangle\,.
\label{eq:PL-symmetric}
\end{align}
This type of backgrounds is called Poisson--Lie symmetric \cite{Klimcik:1995jn}, which sets the ground for Poisson--Lie T-duality and enables us to perform the duality transformation.\footnote{In Poisson--Lie T-duality, $f_{abc}=0$ is assumed originally and then we find $\ell_{am}=0$ and $B^{\text{\tiny WZ}}_{mn}=0$\,. 
The generalized result \eqref{eq:PL-symmetric} has been clearly presented in \cite{Borsato:2021vfy} using the formulation of DFT.
}

To find the Poisson--Lie T-dual background, we choose another maximally isotropic subgroup $\tilde{G}'$, follow the same steps, and then obtain the dual background fields $\{g'_{mn},\,B'_{mn}\}$\,. 
If there are more maximally isotropic subgroups, we can construct more dual backgrounds, and the (classical) equivalence of the sigma models on these backgrounds is called Poisson--Lie T-plurality, or simply Poisson--Lie T-duality.

In a particular case of the Abelian double where $F_{AB}{}^C=0$, by using the parameterization $f=\Exp{x^a\,T_a}$ and $\tilde{g}=\Exp{\tilde{x}_a\,T^a}$, we find $g^f_{ab}=\hat{g}_{ab}$, $B^f_{ab}=\hat{B}_{ab}$, $\ell^a= \rmd x^a$, $\ell_a=0$, and $\tilde{r}_a= \rmd \tilde{x}_a$\,, 
where $\hat{g}_{ab}$ and $\hat{B}_{ab}$ are introduced in Eq.~\eqref{eq:cH-param}, and we obtain a constant background. 
In this case, 
Eq.~\eqref{eq:r-tilde} reduces to
\begin{align}
 \rmd\tilde{x}_a = \hat{B}_{ab}\,\rmd x^b + \hat{g}_{ab}*\rmd x^b \,.
\label{eq:xtildex}
\end{align}
Defining
\begin{align}
 x_L^a = \tfrac{1}{2}\,(\delta^a_b - \hat{g}^{ac}\,\hat{B}_{cb})\,x^b + \tfrac{1}{2}\,\hat{g}^{ab}\,\tilde{x}_b\,,\qquad 
 x_R^a = \tfrac{1}{2}\,(\delta^a_b + \hat{g}^{ac}\,\hat{B}_{cb})\,x^b - \tfrac{1}{2}\,\hat{g}^{ab}\,\tilde{x}_b \,,
\end{align}
the following standard equations are reproduced under the equations of motion:
\begin{align}
 \partial_- x_L^a = 0\,,\qquad \partial_+ x_R^a = 0 \,.
\end{align}
Here we have introduced the light-cone coordinates $\sigma^{\pm}\equiv \tau \pm \sigma$\,, for which 
\begin{align}
 \gamma^{ij} = \begin{pmatrix} 0 & -2 \\ -2 & 0 \end{pmatrix},\qquad
 *\rmd\sigma^{\pm}=\pm \rmd\sigma^{\pm} \,,
\label{eq:LC}
\end{align}
with $\gamma^{ij}$ being the inverse of the world-sheet metric.
The self-duality constraint \eqref{eq:SD} can be understood as a non-Abelian generalization of the above relation \cite{Hull:2004in}. 

\subsection{Self-duality of WZNW models}
\label{sec:self-dual-case}
In the following, we consider a particular case where the string sigma models reduce to WZNW models \cite{Klimcik:1996hp}. 
We assume that the $2d$-dimensional Lie algebra $\mathfrak{d}$ contains a $d$-dimensional subalgebra $\mathfrak{r}$ generated by $t_{\hat{\gga}}$ ($\hat{\gga}=\hat{1},\dotsc,\hat{d}$) satisfying
\begin{align}
 [t_{\hat{\gga}},\,t_{\hat{\ggb}}] = f_{\hat{\gga}\hat{\ggb}}{}^{\hat{\ggc}}\,t_{\hat{\ggc}}\,,\qquad
 \hat{\cH}(t_{\hat{\gga}}) = \pm t_{\hat{\gga}}\,,
 \label{eq:thata}
\end{align}
where either sign can be chosen. 
Using a parameterization\footnote{
In our conventions, $(\hat{g}_{ac}+\hat{B}_{ac})\,(\tilde{g}^{cb}+\beta^{cb})=\delta_a^b$ is satisfied.
}
\begin{align}
 \hat{\cH}(T_A) = \hat{\cH}_{A}{}^B\,T_B = \begin{pmatrix} -(\tilde{g}^{-1}\,\beta)_a{}^b & (\tilde{g}^{-1})_{ab} \\ (\tilde{g}-\beta\,\tilde{g}^{-1}\,\beta)^{ab} & (\beta\,\tilde{g}^{-1})^a{}_b \end{pmatrix} T_B\,,
\end{align}
the second condition indicates that $t_{\hat{\gga}}$ can be expressed as
\begin{align}
 t_{\hat{\gga}} = d_{\hat{\gga}b} \,\bigl(T^{b} \pm E_\mp^{bc}\,T_c\bigr)\qquad \bigl(E_\pm^{ab}\equiv \tilde{g}^{ab}\pm \beta^{ab}\bigr)\,,
\label{eq:t_a-def}
\end{align}
where $d_{\hat{\gga}b}$ is an arbitrary invertible matrix. 

We also assume that the group element $l(\sigma)\in\cD$ can be parameterized as
\begin{align}
 l(\sigma) = f(\sigma)\,\tilde{g}(\sigma) \,,
 \label{eq:lfg}
\end{align}
where $f(\sigma)\in R=\exp(\mathfrak{r})$ and $\tilde{g}\in \tilde{G}$\,.
Since $f\,t_{\hat{\gga}}\,f^{-1}$ is an element of $\mathfrak{r}$\,, we find
\begin{align}
 \hat{\cH}(f\,t_{\hat{\gga}}\,f^{-1}) = \pm f\,t_{\hat{\gga}}\,f^{-1} = f\,\hat{\cH}(t_{\hat{\gga}})\,f^{-1}\qquad (\forall f\in R)\,.
\end{align}
This can be also expressed as
\begin{align}
 \hat{\cH}^f = \hat{\cH}\,.
\label{eq:cH-inv}
\end{align}
In addition, 
we find that $\ell$ satisfies
\begin{align}
 (\hat{\cH}^f)^A{}_B\,\ell^B = \hat{\cH}^A{}_B\,\ell^B = \pm \ell^A\,,
\label{eq:H-ell}
\end{align}
and this can be expressed as
\begin{align}
 \ell_b = \pm g^f_{ab}\,\ell^a - B^f_{ab}\,\ell^a = \pm \hat{g}_{ab}\,\ell^a - \hat{B}_{ab}\,\ell^a \,.
\end{align}
Then the action in \eqref{eq:SM-action22} becomes
\begin{align}
 S &= \frac{1}{4\pi\tilde{\alpha}'}\int_{\Sigma} g_{ab}\,\ell^a \wedge *\ell^b 
 -\frac{1}{24\pi\tilde{\alpha}'}\int_{\cB} \langle \ell ,\, [\ell ,\, \ell]\rangle
\nn\\
 &= \pm \frac{1}{8\pi\tilde{\alpha}'} \int_\Sigma \langle \ell , *\ell \rangle - \frac{1}{24\pi\tilde{\alpha}'}\int_{\cB} \langle \ell , [\ell,\,\ell]\rangle 
\nn\\
 &= \frac{1}{8\pi\tilde{\alpha}'} \int_\Sigma \hh_{\hat{\gga}\hat{\ggb}}\, \ell^{\hat{\gga}} \wedge *\ell^{\hat{\ggb}} \mp \frac{1}{24\pi\tilde{\alpha}'}\int_{\cB} f_{\hat{\gga}\hat{\ggb}\hat{\ggc}}\,\ell^{\hat{\gga}}\wedge \ell^{\hat{\ggb}}\wedge \ell^{\hat{\ggc}} \,,
 \label{eq:S'}
\end{align}
where
\begin{align}
 \hh_{\hat{\gga}\hat{\ggb}}\equiv \bigl\langle t_{\hat{\gga}},\,\hat{\cH}(t_{\hat{\ggb}})\bigr\rangle = 2\,d_{\hat{\gga}c} \,d_{\hat{\ggb}d}\,\tilde{g}^{cd}\,,\qquad 
 f_{\hat{\gga}\hat{\ggb}\hat{\ggc}}\equiv f_{\hat{\gga}\hat{\ggb}}{}^{\hat{\ggd}}\,\hh_{\hat{\ggd}\hat{\ggc}} = f_{[\hat{\gga}\hat{\ggb}\hat{\ggc}]}\,.
\end{align}
This is the WZNW action associated with $R$, as claimed at the beginning of this section.
Changing the parameterization as
\begin{align}
 l(\sigma) = f'(\sigma)\,\tilde{g}'(\sigma) \qquad \bigl(f'(\sigma)\in R\,,\quad \tilde{g}'\in \tilde{G}'\bigr)\,,
 \label{eq:lf'g'}
\end{align}
we obtain the dual action, which is again the WZNW action for the same group $R$. 
In this sense, Poisson--Lie T-duality is self-dual in this case. 

The self-duality constraint \eqref{eq:r-tilde} now reduces to
\begin{align}
 \tilde{r}_a = - \ell_a + \hat{B}_{ab}\,\ell^b + \hat{g}_{ab}*\ell^b = \hat{g}_{ab}\,\bigl(*\ell^b \mp \ell^b\bigr) \,,
\end{align}
which can be expressed as follows in the light-cone coordinates \eqref{eq:LC}:
\begin{align}
 \tilde{r}_{\pm a} = 0 \,,\qquad 
 \tilde{r}_{\mp a} = \mp 2\,\hat{g}_{ab}\,\ell^b_{\mp} \,.
\label{eq:r-tilde2}
\end{align}
Namely, under the equations of motion, $\tilde{g}\in \tilde{G}$ depends only on $\sigma^\mp$\,; $\tilde{g}=\tilde{g}(\sigma^\mp)$\,.

We remark that the self-duality of WZNW models discussed above does not mean that the duality is trivial \cite{Klimcik:1996hp}: 
The duality generally involves non-linear and non-local transformations, and hence the correspondence of the fields and the global properties of the duality are not clear a priori.
We address these issues in the following sections.

\subsection{Current algebra}
\label{sec:current-alg}

The string sigma model on a Poisson--Lie symmetric background can be described by the so-called $\cE$-model \cite{Klimcik:2015gba}. 
This formulation simplifies our discussion on the currents below. 
The $\cE$-model is defined by the Hamiltonian
\begin{align}
 H = \frac{1}{4\pi\tilde{\alpha}'}\int\rmd \sigma\,\hat{\cH}^{AB}\,j_A(\sigma)\,j_B(\sigma)\,,
\label{eq:Ham}
\end{align}
where the currents $j_A(\sigma)$ satisfy the following algebra under the canonical Poisson bracket:
\begin{align}
 \{j_A(\sigma),\,j_B(\sigma')\} = 2\pi\tilde{\alpha}'\,F_{AB}{}^C\,j_C(\sigma)\,\delta(\sigma-\sigma') + 2\pi\tilde{\alpha}'\,\eta_{AB}\,\delta'(\sigma-\sigma')\,.
\label{eq:E-algebra}
\end{align}
The time evolution of the currents is determined by
\begin{align}
 \partial_\tau j_A(\sigma) = \{j_A(\sigma),\,H\}\,.
\label{eq:E-eom}
\end{align}
This equation corresponds to the equations of motion of the string theory on the Poisson--Lie symmetric background. 

The currents $j_A(\sigma)$ of the $\cE$-model can be constructed as
\begin{align}
 j_A(\sigma) = E_A{}^M\,Z_M\,,\qquad Z_M \equiv \begin{pmatrix} 2\pi\tilde{\alpha}'\,p_m(\sigma) \\ \partial_\sigma x^m(\sigma) \end{pmatrix} ,
\end{align}
where $p_m$ are the canonical conjugate momenta of $x^m$ satisfying $\{x^m(\sigma),\,p_n(\sigma')\}=\delta^m_n\,\delta(\sigma-\sigma')$ and the generalized frame fields $E_A{}^M$ are given by \cite{Borsato:2021vfy}
\begin{align}
 E_A{}^M = M_A{}^B\,\hat{V}_B{}^N\,N_N{}^M\,,\quad 
 \hat{V}_A{}^M \equiv \begin{pmatrix} v_a^m & 0 \\ 0 & \ell^a_m \end{pmatrix},\quad
 N_M{}^N \equiv \begin{pmatrix} \delta_m^n & -(B^{\text{\tiny WZ}}_{mn} + \ell_{a[m} \ell^a_{n]}) \\ 0 & \delta^m_n \end{pmatrix},
\end{align}
where $v_a^m\,\ell^b_m=\delta_a^b$\,. 
They satisfy the algebra \cite{Hassler:2017yza,Demulder:2018lmj,Borsato:2021vfy}
\begin{align}
 [E_A,\,E_B]_{\text{D}}^M = - F_{AB}{}^C\,E_C{}^M\,,
\end{align}
where $[\cdot ,\,\cdot]_{\text{D}}$ is called the D-bracket\footnote{
For two generalized vector fields $V^M=(v^m,\,\tilde{v}_m)$ and $W^M=(w^m,\,\tilde{w}_m)$ which depend on $x^m$\,, the D-bracket is defined as $[V,\,W]_{\text{D}}^M = (\pounds_v w^m,\, [\pounds_v \tilde{w} - \iota_w\rmd \tilde{v}]_m)$\,.
} and $F_{AB}{}^C$ are the structure constants of the Lie algebra $\mathfrak{d}$\,. 
Using this property and the canonical commutation relation of $x^m(\sigma)$ and $p_m(\sigma)$, we can reproduce the current algebra \eqref{eq:E-algebra} \cite{Siegel:1993th}. 

The generalized frame fields are constructed from the coset element $f\in \cD/\tilde{G}$, 
and hence so are the currents $j_A(\sigma)$.
To consider the dual model, let us choose another maximally isotropic subgroup $\tilde{G}'$.
We can construct new currents $j'_A(\sigma)$ from the new coset $\cD/\tilde{G}'$\,. 
We here use a convention where the generators $T_A$ are redefined as $T'_A=C_A{}^B\,T_B$ with $C_A{}^B \in$ O($d,d$) \cite{Snobl:2002kq} such that $\langle T'_A,\,T'_B\rangle=\eta_{AB}$ is satisfied and the generators $T'^a$ form the Lie algebra $\tilde{\mathfrak{g}}'$ associated with the Lie group $\tilde{G}'$. 
The structure constants of the new generators become $F'_{AB}{}^C=C_A{}^D\,C_B{}^E\,(C^{-1})_F{}^C\,F_{DE}{}^F$ and the currents constructed from $\cD/\tilde{G}'$ satisfy
\begin{align}
 \{j'_A(\sigma),\,j'_B(\sigma')\} = 2\pi\tilde{\alpha}'\,F'_{AB}{}^C\,j'_C(\sigma)\,\delta(\sigma-\sigma') + 2\pi\tilde{\alpha}'\,\eta_{AB}\,\delta'(\sigma-\sigma')\,.
\end{align}
If we go back to the original basis by defining $\tilde{j}_A(\sigma)\equiv (C^{-1})_A{}^B\,j'_B(\sigma)$, however, $\tilde{j}_A$ satisfy the algebra of the same form as \eqref{eq:E-algebra}. 
By defining the Hamiltonian of the dual model as
\begin{align}
 \tilde{H} = \frac{1}{4\pi\tilde{\alpha}'}\int\rmd \sigma\,\hat{\cH}^{AB}\,\tilde{j}_A(\sigma)\,\tilde{j}_B(\sigma)\,,
\end{align}
the equations of motion \eqref{eq:E-eom} then take the same form in terms of $\tilde{j}_A(\sigma)$,
\begin{align}
 \partial_\tau \tilde{j}_A(\sigma) = \{\tilde{j}_A(\sigma),\,\tilde{H}\}\,,
\end{align}
which shows the (classical) equivalence between the original and the dual model. 
The canonical equivalence at the classical level has also been shown in \cite{Klimcik:1995dy,Sfetsos:1997pi}.
In particular, the generating functionals of the canonical transformations have been constructed in the latter reference. 

\subsection{Current algebra of WZNW models}
\label{sec:CA-WZW}
Now we shall consider the case discussed in section \ref{sec:self-dual-case}, and find the relation between the current algebra of $j_A(\sigma)$ and the more familiar current algebra of WZNW models. 
Corresponding to the definition of $t_{\hat{\gga}}$ in Eq.~\eqref{eq:t_a-def}, if we define the generalized vector fields
\begin{align}
 V_{\hat{\gga}}{}^M \equiv d_{\hat{\gga}b} \,\bigl(E^{b M} \pm E_\mp^{bc}\,E_c{}^M\bigr)\,,
\end{align}
they satisfy the Lie algebra $\mathfrak{r}$,
\begin{align}
 [V_{\hat{\gga}},\,V_{\hat{\ggb}}]_{\text{D}}^M = - f_{\hat{\gga}\hat{\ggb}}{}^{\hat{\ggc}}\,V_{\hat{\ggc}}{}^M\,.
\label{eq:V-alg}
\end{align}
Moreover, due to the property \eqref{eq:cH-inv}, they are generalized Killing vectors satisfying
\begin{align}
 \hat{\pounds}_{V_{\hat{\gga}}} \cH_{MN}\equiv V^P\,\partial_P\cH_{MN} + 2\,\bigl(\partial_{(M} V^P - \partial^P V_{(M}\bigr)\,\cH_{N)P}=0\,.
\end{align}
Namely, there exist the corresponding (on-shell) conserved currents
\begin{align}
 \cJ_{\hat{\gga}} = \gamma\,\bigl[V_{\hat{\gga}}{}^m\,\bigl(g_{mn}\,*\rmd x^n + B_{mn}\,\rmd x^n\bigr) + V_{\hat{\gga}}{}_n\,\rmd x^n \bigr] \,,
 \label{eq:cJa}
\end{align}
where $\gamma$ is a normalization constant. 
Using $\cJ_{\hat{\gga}\sigma}=\gamma\,V_{\hat{\gga}}{}^M\,Z_{M}$ and Eq.~\eqref{eq:E-algebra}, we can easily compute the Poisson bracket of the spatial components $\cJ_{\hat{\gga}\sigma}(\sigma)$ of \eqref{eq:cJa} as
\begin{align}
 \{\cJ_{\hat{\gga}\sigma}(\sigma),\,\cJ_{\hat{\ggb}\sigma}(\sigma')\} = 2\pi\tilde{\alpha}'\,\gamma\,f_{\hat{\gga}\hat{\ggb}}{}^{\hat{\ggc}}\,\cJ_{\hat{\ggc}\sigma}(\sigma)\,\delta(\sigma-\sigma') \pm 2\pi\tilde{\alpha}'\,\gamma^2\,\hh_{\hat{\gga}\hat{\ggb}}\,\delta'(\sigma-\sigma')\,.
\end{align}
In the light-cone coordinates, the components of the currents are
\begin{align}
 \cJ_{\hat{\gga}\pm} = \gamma\,\bigl[V_{\hat{\gga}}{}^m\,\bigl(g_{mn}\pm B_{mn}\bigr) \pm V_{\hat{\gga}}{}_n\bigr]\,\partial_\pm x^n\,.
\label{eq:J}
\end{align}
Due to the property $\cH^M{}_N\,V_{\hat{\gga}}{}^N= \pm V_{\hat{\gga}}{}^M$ that follows from $\hat{\cH}(t_{\hat{\gga}}) = \pm t_{\hat{\gga}}$, we have $\cJ_{\hat{\gga}\mp}=0$ for $\hat{\cH}(t_{\hat{\gga}}) = \pm t_{\hat{\gga}}$\,. 
Then, recalling $\cJ_{\hat{\gga}\sigma}= \cJ_{\hat{\gga}+} - \cJ_{\hat{\gga}-} = \pm \cJ_{\hat{\gga}\pm}$\,, we find
\begin{align}
 \{\cJ_{\hat{\gga}\pm}(\sigma),\,\cJ_{\hat{\ggb}\pm}(\sigma')\} = \pm 2\pi\tilde{\alpha}'\,\gamma \,f_{\hat{\gga}\hat{\ggb}}{}^{\hat{\ggc}}\,\cJ_{\hat{\ggc}\pm}(\sigma)\,\delta(\sigma-\sigma') \pm 2\pi\tilde{\alpha}'\,\gamma^2\,\hh_{\hat{\gga}\hat{\ggb}}\,\delta'(\sigma-\sigma')\,,
\end{align}
or equivalently, using the conservation of the currents $\partial_\mp \cJ_{\hat{\gga}\pm}=0$\,, 
\begin{align}
 \{\cJ_{\hat{\gga}\pm}(\sigma^\pm),\,\cJ_{\hat{\ggb}\pm}(\sigma^{\prime\pm})\} = \pm 2\pi\tilde{\alpha}'\,\gamma \,f_{\hat{\gga}\hat{\ggb}}{}^{\hat{\ggc}}\,\cJ_{\hat{\ggc}\pm}(\sigma^\pm)\,\delta(\sigma^\pm-\sigma^{'\pm}) + 2\pi\tilde{\alpha}'\,\gamma^2\,\hh_{\hat{\gga}\hat{\ggb}}\,\delta'(\sigma^{\pm}-\sigma^{\prime\pm})\,.
\end{align}
This corresponds to the current algebra of the left or the right currents in the WZNW model, depending on the choice of the sign in \eqref{eq:thata}.
Since either set of $\cJ_{\hat{\gga}-}$ or $\cJ_{\hat{\gga}+}$ vanishes, we need to elaborate 
a little more to obtain both left and right currents.

For this purpose, we define the generalized vectors of opposite chirality
\begin{align}
 K_{\hat{\gga}}{}^M \equiv - d_{\hat{\gga}b} \,\bigl(E^{b M} \mp E_\pm^{bc}\,E_c{}^M\bigr)\,,
\label{eq:K-def}
\end{align}
which satisfy $\cH^M{}_N\,K_{\hat{\gga}}{}^N= \mp K_{\hat{\gga}}{}^M$.
Defining the corresponding currents by
\begin{align}
 \cJ'_{\hat{\gga}\mp} = \gamma\,\bigl[K_{\hat{\gga}}{}^m\,\bigl(g_{mn}\mp B_{mn}\bigr) \mp K_{\hat{\gga}}{}_n\bigr]\,\partial_\mp x^n\,,
\label{eq:cJp}
\end{align}
the Hamiltonian \eqref{eq:Ham} can be expressed as
\begin{align}
 H = \frac{1}{4\pi\tilde{\alpha}'\gamma^2}\,\hh^{\hat{\gga}\hat{\ggb}}\,\bigl(\cJ_{\hat{\gga}\pm}\,\cJ_{\hat{\ggb}\pm} + \cJ'_{\hat{\gga}\mp}\,\cJ'_{\hat{\ggb}\mp}\bigr)\,,
\label{eq:Hamiltonian}
\end{align}
where $\hh^{\hat{\gga}\hat{\ggc}}\,\hh_{\hat{\ggc}\hat{\ggb}}=\delta^{\hat{\gga}}_{\hat{\ggb}}$ and we have used
\begin{align}
 \hh^{\hat{\gga}\hat{\ggb}}\,\cJ_{\hat{\gga}\pm}\,\cJ_{\hat{\ggb}\pm}
 = \frac{\gamma^2}{2}\,\tilde{g}_{ab}\,(j^a \pm E^{ac}_\mp\,j_c) \,(j^b \pm E^{bd}_\mp\,j_d) 
 = \frac{\gamma^2}{2}\,\bigl(\hat{\cH}_{AB}\pm \eta_{AB}\bigr)\,j^A\,j^B \,,
\end{align}
and a similar expression for $\cJ'_{\hat{\gga}\mp}$\,. 
Then one might expect that $\cJ'_{\hat{\gga}\mp}$ would correspond to the right/left currents in the WZNW model. 
This is, however, not always the case because $K_{\hat{\gga}}{}^M$ are not generalized Killing vectors in general, and moreover, they do not satisfy the algebra similar to \eqref{eq:V-alg}.\footnote{
There is a specific class of Drinfel'd doubles where $\cJ'_{\hat{\gga}\mp}$ become the right/left currents in the WZNW model (see in Appendix \ref{app:lambda} for the details).
} 
In fact, we need to find a suitable coordinate-dependent redefinition\footnote{
A similar redefinition has been discussed in \cite{Blumenhagen:2015zma}.
}
\begin{align}
 \bar{V}_{\hat{\gga}}{}^M \equiv \Lambda_{\hat{\gga}}{}^{{\hat{\ggb}}}\,K_{\hat{\ggb}}{}^M\qquad \bigl(\Lambda_{\hat{\gga}}{}^{{\hat{\ggc}}}\,\Lambda_{\hat{\ggb}}{}^{{\hat{\ggd}}}\,\hh_{\hat{\ggc}\hat{\ggd}}=\hh_{\hat{\gga}\hat{\ggb}}\bigr)\,,
\end{align}
such that the resulting $\bar{V}_{\hat{\gga}}{}^M$ satisfy
\begin{align}
 [\bar{V}_{\hat{\gga}},\,\bar{V}_{\hat{\ggb}}]_{\text{D}}^M = + f_{\hat{\gga}\hat{\ggb}}{}^{\hat{\ggc}}\,\bar{V}_{\hat{\ggc}}{}^M\,,\qquad
 \hat{\pounds}_{\bar{V}_{\hat{\gga}}} \cH_{MN}=0\,.
\end{align}
We then define the corresponding conserved currents as
\begin{align}
 \bar{\cJ}_{\hat{\gga}\mp} = \Lambda_{\hat{\gga}}{}^{{\hat{\ggb}}} \,\cJ'_{\hat{\ggb}\mp} \,,
\label{eq:Jtilde}
\end{align}
which satisfy 
$\bar{\cJ}_{\hat{\gga}\pm}=0$ for $\hat{\cH}(t_{\hat{\gga}}) = \pm t_{\hat{\gga}}$\,. 
These currents turn out to be the right/left currents in the WZNW model satisfying
\begin{align}
 \{\bar{\cJ}_{\hat{\gga}\mp}(\sigma^\mp),\,\bar{\cJ}_{\hat{\ggb}\mp}(\sigma^{\prime\mp})\} = \pm 2\pi\tilde{\alpha}'\,\gamma \,f_{\hat{\gga}\hat{\ggb}}{}^{\hat{\ggc}}\,\bar{\cJ}_{\hat{\ggc}\mp}(\sigma^{\mp})\,\delta(\sigma^\mp-\sigma^{\prime\mp}) + 2\pi\tilde{\alpha}'\,\gamma^2\,\hh_{\hat{\gga}\hat{\ggb}}\,\delta'(\sigma^\mp-\sigma^{\prime\mp})\,.
\end{align}
Since this $\cE$-model is the WZNW model, this matrix $\Lambda_{\hat{\gga}}{}^{{\hat{\ggb}}}$ should exist, and indeed can be found in the examples studied in this paper. 

In section \ref{sec:PLWZW} and later, we choose the convention $\hat{\cH}(t_{\hat{\gga}}) = + t_{\hat{\gga}}$ and use the short-hand notation $\cJ_{\hat{\gga}} \equiv \cJ_{\hat{\gga}+}$ and $\bar{\cJ}_{\hat{\gga}} \equiv \bar{\cJ}_{\hat{\gga}-}$\,. 
Then, starting from the currents $j_A$ of the $\cE$-model, we obtain the left and right current algebras
\begin{align}
\begin{split}
 \{\cJ_{\hat{\gga}}(\sigma^+),\,\cJ_{\hat{\ggb}}(\sigma^{\prime +})\} &= 2\pi\tilde{\alpha}'\,\gamma \,f_{\hat{\gga}\hat{\ggb}}{}^{\hat{\ggc}}\,\cJ_{\hat{\ggc}}(\sigma^+)\,\delta(\sigma^+-\sigma^{\prime +}) + 2\pi\tilde{\alpha}'\,\gamma^2\,\hh_{\hat{\gga}\hat{\ggb}}\,\delta'(\sigma^+-\sigma^{\prime +})\,,
\\
 \{\bar{\cJ}_{\hat{\gga}}(\sigma^-),\,\bar{\cJ}_{\hat{\ggb}}(\sigma^{\prime -})\} &= 2\pi\tilde{\alpha}'\,\gamma \,f_{\hat{\gga}\hat{\ggb}}{}^{\hat{\ggc}}\,\bar{\cJ}_{\hat{\ggc}}(\sigma^-)\,\delta(\sigma^--\sigma^{\prime -}) + 2\pi\tilde{\alpha}'\,\gamma^2\,\hh_{\hat{\gga}\hat{\ggb}}\,\delta'(\sigma^--\sigma^{\prime -})\,.
\end{split}
\label{eq:J-alg}
\end{align}
The Hamiltonian \eqref{eq:Hamiltonian} can be expressed as
\begin{align}
 H = - \frac{1}{2\pi}\,(T_{++}+T_{--}) \,, \quad
 T_{++} \equiv -\frac{1}{2\tilde{\alpha}'\gamma^2}\,\hh^{\hat{\gga}\hat{\ggb}}\, \cJ_{\hat{\gga}}\,\cJ_{\hat{\ggb}} \,,\quad
 T_{--} \equiv -\frac{1}{2\tilde{\alpha}'\gamma^2}\,\hh^{\hat{\gga}\hat{\ggb}}\, \bar{\cJ}_{\hat{\gga}}\,\bar{\cJ}_{\hat{\ggb}} \,.
\label{eq:TTbar-def}
\end{align}

\subsection{Conformal invariance in the self-duality of WZNW models}
\label{sec:dilaton}

So far, we have focused on the metric and the $B$-field. 
To describe strings, we further need to take into account the dilaton.
In this respect, some issues have been discussed in the literature regarding non-Abelian T-duality for isometries with $f_{ab}{}^b\neq 0$ \cite{Gasperini:1993nz,Giveon:1993ai,Alvarez:1994np,Elitzur:1994ri} or Poisson--Lie T-duality with $f_b{}^{ba}\neq 0$ \cite{Tyurin:1995bu,VonUnge:2002xjf}.
Here, assuming that $\hat{E}_\pm^{ab}$ is non-degenerate, we show that these issues do not arise in the case of the self-duality of WZNW models. 
We shall employ the flux formulation of DFT \cite{Geissbuhler:2013uka} for this purpose.

In the flux formulation, using the generalized frame fields $E_A{}^M$, we define two fluxes,
\begin{align}
 F_{AB}{}^C \equiv - [E_A,\,E_B]_{\text{D}}^M\,E_M{}^C\,,\qquad 
 F_A \equiv 2\, D_A \mathsf{d} - \partial_M E_A{}^M \,,
\end{align}
where $D_A\equiv E_A{}^M\,\partial_M$ and $\mathsf{d}$ is the T-duality-invariant dilaton of DFT, which is related to the standard dilaton as $\Exp{-2\mathsf{d}}=\Exp{-2\phi}\sqrt{\lvert \det(g_{mn})\rvert}$\,. 
We assume that these fluxes satisfy the consistency condition (or the section condition) \cite{Geissbuhler:2013uka}
\begin{align}
 \cZ\equiv D^AF_A -\tfrac{1}{2}\,F_A\,F^A + \tfrac{1}{12}\,F_{ABC}\,F^{ABC} &= 0\,,
\label{eq:SC1}
\\
 \cZ_{AB}\equiv D_AF_B -D_BF_A + (F_C-D_C)F^C{}_{AB} &= 0\,.
\label{eq:SC2}
\end{align}
In our set-up, $F_{AB}{}^C$ coincide with the structure constants of the Lie algebra $\mathfrak{d}$, and thus are constant. 
For simplicity, we ignore the existence of spectator fields, or focus on the $d$-dimensional space.
The supergravity equations of motion, or equivalently the conditions for the vanishing $\beta$-functions, are then given by
\begin{align}
 \cR=0\,,\qquad 
 R^+_{AB} = 0\quad \bigl(\text{or }R^-_{AB} = 0\bigr)\,,
\end{align}
where
\begin{align}
\begin{split}
 \cR &= 2\,P^{AB}\, \bigl(2\,D_{A} F_{B} - F_{A}\,F_{B}\bigr) + \tfrac{1}{3}\,F_{\ubar{A}\ubar{B}\ubar{C}}\,F^{\ubar{A}\ubar{B}\ubar{C}}
 + F_{\ubar{A}\ubar{B}\bar{C}}\,F^{\ubar{A}\ubar{B}\bar{C}} 
\\
 &= -2\,\bar{P}^{AB}\, \bigl(2\,D_{A} F_{B} - F_{A}\,F_{B}\bigr) - \tfrac{1}{3}\,F_{\bar{A}\bar{B}\bar{C}}\,F^{\bar{A}\bar{B}\bar{C}}
 - F_{\bar{A}\bar{B}\ubar{C}}\,F^{\bar{A}\bar{B}\ubar{C}}\,,
\end{split}
\\
 R^+_{AB}
 &= - F_{\ubar{C}}\,F_{\ubar{A}\bar{B}}{}^{\ubar{C}}
 + D_{\bar{B}}F_{\ubar{A}}
 + F_{\ubar{A}\ubar{C}\bar{D}}\,F_{\bar{B}}{}^{\ubar{C}\bar{D}}\,,
\\
 R^-_{AB}
 &= - F_{\bar{C}}\, F_{\bar{A}\ubar{B}}{}^{\bar{C}}
 + D_{\ubar{B}}F_{\bar{A}}
 + F_{\bar{A}\bar{C}\ubar{D}}\,F_{\ubar{B}}{}^{\bar{C}\ubar{D}}\,,
\end{align}
and 
$R^+_{AB}=0$ is equivalent to $R^-_{AB}=0$ under Eq.~\eqref{eq:SC2}. 
Here we have defined the projectors
\begin{align}
 P^{AB} \equiv \frac{1}{2}\,\bigl(\eta^{AB}+\hat{\cH}^{AB}\bigr)\,,\qquad
 \bar{P}^{AB} \equiv \frac{1}{2}\,\bigl(\eta^{AB}-\hat{\cH}^{AB}\bigr)\,,
\end{align}
and used the notation such as $F_{\ubar{A}}=P_A{}^B\,F_{B}$ or $F_{\bar{A}}=\bar{P}_A{}^B\,F_{B}$\,. 
We note that using the parameterization
\begin{align}
 \cH_{MN} = \begin{pmatrix} g_{mn}-B_{mp}\,g^{pq}\,B_{qn} & B_{mp}\,g^{pn} \\ -g^{mp}\,B_{pn} & g^{mn} \end{pmatrix}
 = E_M{}^A\,E_N{}^B\,\hat{\cH}_{AB}\,,
\end{align}
the generalized Ricci scalar $\cR$ is expressed as follows in terms of the supergravity fields:
\begin{align}
 \cR = R + 4\,\nabla^m \partial_m \phi - 4\,g^{mn}\,\partial_m \phi\,\partial_n \phi - \tfrac{1}{12}\,H_{mnp}\,H^{mnp}\,.
\end{align}

When we study Poisson--Lie T-duality, the dilaton is determined in such a way that $F_A=0$ \cite{Hassler:2017yza,Demulder:2018lmj}.\footnote{
A more general case with $F_A$ is a non-zero constant has been studied, for example, in \cite{Sakatani:2019jgu}.
}
Parameterizing the dilaton $\Exp{-2\mathsf{d}}$ as
\begin{align}
 \Exp{-2\mathsf{d}} = \Exp{-2\hat{\mathsf{d}}} \,\lvert\det(\ell_m^a)\rvert \,,
\end{align}
it is also possible to compute the curved components of the flux $F_M\equiv E_M{}^A\,F_A$ as
\begin{align}
 F_M = 2\, \partial_M \hat{\mathsf{d}} + f_b{}^{ba} \begin{pmatrix}
 \ell_{am}+ v_a^n\, (B^{\text{\tiny WZ}}_{nm} + \ell_{a[n}\,\ell^a_{m]}) \\
 -v_a^m
\end{pmatrix}.
\label{eq:F-general}
\end{align}
Then, if the dual structure constants satisfy $f_b{}^{ba}=0$\,, 
we can realize $F_A=0$ by choosing $\hat{\mathsf{d}}$ as a constant. 
However, the condition $f_b{}^{ba}=0$ is not $\OO(d,d)$ covariant, 
and thus it can be broken in the dual frame even when it holds in the original duality frame.
Such an example has been found in \cite{Gasperini:1993nz} where the dilaton cannot be found such that the supergravity equations of motion are satisfied. 
Namely, the conformal invariance can be broken when $f_b{}^{ba}\neq 0$ in the dual model, and this was understood in terms of a mixed gauge and gravitational anomaly in \cite{Giveon:1993ai,Alvarez:1994np,Elitzur:1994ri}.\footnote{
Though the example in \cite{Gasperini:1993nz} is for non-Abelian T-duality, the argument generally holds also for Poisson--Lie T-duality \cite{Tyurin:1995bu}. 
Poisson--Lie T-duality reduces to non-Abelian T-duality e.g.~when $f_a{}^{bc} =0$.
} 

Accordingly, from a conservative point of view, it may not be appropriate to consider Poisson--Lie T-duality when the original or the dual model contains non-vanishing $f_b{}^{ba}$\,. 
However, even in such a case, the duality transformation works as a solution generating transformation in DFT. 
Indeed, even in the problematic example of \cite{Gasperini:1993nz}, the dual background has been identified as a solution of DFT in \cite{Fernandez-Melgarejo:2017oyu}. 
Since DFT is a framework providing useful insights into the low-energy effective theory of strings, this result suggests that there might be a resolution to these issues (see also \cite{Sakamoto:2017wor,Fernandez-Melgarejo:2018wpg}).
A common understanding however is still absent. 

In the case of the self-duality of WZNW models, the situation is simpler even though $f_b{}^{ba}\neq 0$ in general.
Indeed, in the example studied in section \ref{sec:ex1}, both the original and dual model satisfies $f_b{}^{ba}\neq 0$\,. 
However, as we show below, this does not matter because the flux $F_A$ is decoupled from the supergravity equations of motion and the conformal invariance can be realized both in the original and the dual model.
A similar situation has been observed in the context of the Yang--Baxter deformation in \cite{Sakamoto:2018krs}. 
A more general discussion associated with the decoupling has been given in \cite{Wulff:2018aku}.\footnote{
A field redefinition that cancels the mixed anomaly associated with $f_{ab}{}^b\neq 0$ has also been discussed.
} 
It would be interesting to find a unified understanding.

As shown in Eqs.~\eqref{eq:cH-inv} and \eqref{eq:H-ell}, we have
\begin{align}
 \hat{\cH}^f_{AB}=\hat{\cH}_{AB} \,,\qquad \hat{\cH}_A{}^B\,\ell_B = \pm \ell_A\,,
\end{align}
when $\hat{\cH}(t_{\hat{\gga}}) = \pm t_{\hat{\gga}}$\,. 
The latter gives
\begin{align}
 \ell_{am} = (\pm \hat{g}_{ab} + \hat{B}_{ab})\,\ell^b_m\,,
\end{align}
from which it follows that the former, together with \eqref{eq:PL-symmetric}, results in
\begin{align}
 g_{mn}= \hat{g}_{ab}\,\ell^a_m\,\ell^b_n\,,\qquad
 B_{mn}= B^{\text{\tiny WZ}}_{mn}\,.
\end{align}
Using these, we find that $F_M$ of \eqref{eq:F-general} reduces to
\begin{align}
 F_M = 2\, \partial_M \hat{\mathsf{d}} + f_b{}^{ba} \begin{pmatrix}
 v_a^n\, (\pm g_{nm} + B_{nm}) \\
 -v_a^m
\end{pmatrix},
\end{align}
and by choosing $\hat{\mathsf{d}}$ to be constant, 
\begin{align}
 \hat{\cH}_A{}^B\,F_B = \mp F_A\,.
\end{align}
Consequently, we have $F_{\ubar{A}} = P_A{}^B\,F_{B} = 0$ (or $F_{\bar{A}} = \bar{P}_A{}^B\,F_{B} = 0$) for $\hat{\cH}(t_{\hat{\gga}}) = + t_{\hat{\gga}}$ (or $-t_{\hat{\gga}}$)\,. 
Recalling $F_{\ubar{A}\ubar{C}\bar{D}}=0$ (or $F_{\bar{A}\bar{C}\ubar{D}}=0$), which follows from $[t_{\hat{\gga}},\,t_{\hat{\ggb}}] = f_{\hat{\gga}\hat{\ggb}}{}^{\hat{\ggc}}\,t_{\hat{\ggc}}$, we then find $R^+_{AB}=0$ (or $R^-_{AB}=0$). 
Moreover, since $\cR = \tfrac{1}{3}\,F_{\ubar{A}\ubar{B}\ubar{C}}\,F^{\ubar{A}\ubar{B}\ubar{C}}$ (or $\cR= - \tfrac{1}{3}\,F_{\bar{A}\bar{B}\bar{C}}\,F^{\bar{A}\bar{B}\bar{C}}$), 
$F_{A}$ does not give any contribution to the supergravity equations of motion even when it is not constant.
In terms of the structure constants $f_{\hat{\gga}\hat{\ggb}}{}^{\hat{\ggc}}$, the above result is written as $\cR = \tfrac{1}{3}\,f_{\hat{\gga}\hat{\ggb}\hat{\ggc}}\,f^{\hat{\gga}\hat{\ggb}\hat{\ggc}}$ where the indices are raised/lowered with the metric $\hh_{\hat{\gga}\hat{\ggb}}$ and its inverse. 

To summarize, from a given $2d$-dimensional algebra $\mathfrak{d}$ and the constant matrix $\hat{\cH}_{AB}$\,, we can construct the supergravity fields as
\begin{align}
 g_{mn}= \hat{g}_{ab}\,\ell^a_m\,\ell^b_n\,\bigl(=2\,\hh_{\hat{\gga}\hat{\ggb}}\,\ell_m^{\hat{\gga}}\,\ell_n^{\hat{\ggb}}\bigr)\,,\qquad
 B_{mn}= B^{\text{\tiny WZ}}_{mn}\,,\qquad
 \phi = \hat{\phi}\,,
\end{align}
where we have parameterized the constant $\Exp{-2\hat{\mathsf{d}}}$ as $\Exp{-2\hat{\mathsf{d}}}=\Exp{-2\hat{\phi}}\sqrt{\lvert\det(\hat{g}_{ab})\rvert}$\,. 
Then even when $f_b{}^{ba}\neq 0$, the flux $F_A$ does not contribute to the supergravity equations of motion (of the $d$-dimensional part), so that
\begin{align}
 \cR = \tfrac{1}{3}\,f_{\hat{\gga}\hat{\ggb}\hat{\ggc}}\,f^{\hat{\gga}\hat{\ggb}\hat{\ggc}}\,,\qquad 
 R^\pm_{AB} = 0\,.
\end{align}
By adding other spacetime that is orthogonal to this $d$-dimensional space such that the constant term in $\cR$ is canceled out, we obtain a conformal sigma model as usual. 

\section{Poisson--Lie T-duality of SU(2) WZNW model}
\label{sec:PLWZW}
In this section, we find concrete examples of the self-duality of WZNW models, where the Drinfel'd double is six-dimensional and the Lie group $R$ turns out to be SU(2). 
These examples allow us to study their quantum aspects in detail, which are discussed in section \ref{sec:quantum}. 

\subsection{Example 1}
\label{sec:ex1}
Let us take a Drinfel'd double given by\footnote{
Note that this Drinfel'd double satisfies $F_{ABC}\,F^{ABC}\neq 0$ and is of the type called non-geometric in \cite{Dibitetto:2012rk}. 
Indeed, if $F_A=0$, the section condition $\cZ=0$ in \eqref{eq:SC1} is broken. 
However, we are introducing non-constant $F_A$ (note that $f_b{}^{ba}\neq 0$) such that $\cZ=0$, and hence the section condition is not broken in our example.
}
\begin{align}
\begin{split}
 f_{13}{}^1&=-\omega\,,\quad
 f_{13}{}^2=1\,,\quad
 f_{23}{}^1=-1\,,\quad
 f_{23}{}^2=-\omega\,,
\\
 f_1{}^{13} &= -\omega\,,\quad
 f_2{}^{13} = \omega^2\,,\quad
 f_1{}^{23} = -\omega^2\,,\quad
 f_2{}^{23} = -\omega\quad (\omega>0)\,,
\end{split}
\end{align}
the Lie algebra of which is isomorphic to $\mathfrak{sl}(2,\mathbb{C})$. 
In the classification of \cite{Snobl:2002kq}, this corresponds to the Manin triple $(\bm{7_\omega}\vert \bm{7_{1/\omega}}\vert \bm{-\omega^2})$. 
We note that this example has been discussed in \cite{Klimcik:1996hp}. 
Here, we take $ \bm{7_{1/\omega}}$ to be the dual Lie subalgebra generated by $T^a$. The corresponding Lie subgroup is denoted by $\tilde{G}_{1/\omega}$ in the following. 

We choose the constant matrix $E_+^{ab}$ as
\begin{align}
 E_+^{ab} = \begin{pmatrix} 1 & \frac{\omega^2-1}{2\omega} & 0 \\ -\frac{\omega^2-1}{2\omega} & 1 & 0 \\ 0 & 0 & 1 \end{pmatrix},
\label{eq:E-ex1}
\end{align}
and then the generators $t_{\hat{\gga}}\equiv -\frac{1}{1+\omega^2}\,\delta_{\hat{\gga}b}\,\bigl(T^{b} + T_c\,E_+^{cb}\bigr)$, which span $\mathfrak{r}$\,, satisfy the $\mathfrak{su}(2)$ algebra
\begin{align}
 [t_{\hat{1}},\,t_{\hat{2}}] = t_{\hat{3}}\,,\qquad
 [t_{\hat{2}},\,t_{\hat{3}}] = t_{\hat{1}}\,,\qquad
 [t_{\hat{3}},\,t_{\hat{1}}] = t_{\hat{2}}\,.
\end{align} 
From \eqref{eq:S'}, we obtain the SU(2) WZNW model with the action, 
\begin{align}
 S_{\text{\tiny WZNW}}(f)\equiv S = - \frac{k}{8\pi} \int_\Sigma \kappa(\ell, *\ell) + \frac{k}{24\pi}\int_{\cB} \kappa( \ell,\,[\ell,\,\ell]) \,.
\end{align}
We have introduced $k\equiv \frac{4}{(1+\omega^2)^2\,\tilde{\alpha}'}$, which has to be a positive integer, and a bilinear form $\kappa(t_{\hat{\gga}},\,t_{\hat{\ggb}})=-\frac{1}{2}\,\delta_{\hat{\gga}\hat{\ggb}}$\,. 
We shall parameterize $f$ and $\tilde{g}$ as 
\begin{align}
 f \equiv \Exp{-2\,u\,t_{\hat{3}}}\Exp{-2\,\rho\,t_{\hat{1}}}\Exp{-2\,v\,t_{\hat{3}}}\,,\qquad 
 \tilde{g} \equiv \Exp{\frac{\tilde{\rho}}{1+\omega^2} \,T^3}\Exp{\tilde{y} \,T^2}\Exp{\tilde{x} \,T^1}\,,
\end{align}
where the normalization of $\tilde{\rho}$ has been fixed for later convenience.
Then by using
\begin{align}
 \hat{g}_{ab} + \hat{B}_{ab} &= \begin{pmatrix}
 \frac{4 \omega^2}{(1+\omega^2)^2} & \frac{2 \omega (1-\omega^2)}{(1+\omega^2)^2} & 0 \\
 -\frac{2 \omega (1-\omega^2)}{(1+\omega^2)^2} & \frac{4 \omega^2}{(1+\omega^2+1)^2} & 0 \\
 0 & 0 & 1\end{pmatrix},\quad
 B^{\text{\tiny WZ}} = \frac{4\cos 2 \rho}{(1 + \omega^2)^2} \,\rmd u\wedge \rmd v\,,
\\
 \ell_m{}^a &=
\begin{pmatrix}
 \frac{\sin 2\rho\, (2 \omega \sin 2v + (1-\omega^2) \cos 2v)}{\omega(1+\omega^2)} & -\frac{2 \sin 2\rho\,(\omega \cos v -\sin v) (\omega \sin v +\cos v)}{\omega(1+\omega^2)} & \frac{2 \cos 2\rho}{1+\omega^2} \\
 0 & 0 & \frac{2}{1+\omega^2} \\
 \frac{-(1-\omega^2) \sin 2 v +2\omega \cos 2 v}{\omega(1+\omega^2)} & \frac{2 \omega \sin 2v +(1-\omega^2) \cos 2 v}{\omega(1+\omega^2)} & 0\end{pmatrix} ,
\\
 \ell_{ma}&=\begin{pmatrix}
 \frac{2 \sin 2 \rho \sin 2 v}{1+\omega^2} & -\frac{2 \sin 2 \rho \cos 2 v}{1+\omega^2} & \frac{2 \cos 2 \rho}{1+\omega^2} \\
 0 & 0 & \frac{2}{1+\omega^2} \\
 \frac{2 \cos 2 v}{1+\omega^2} & \frac{2 \sin 2 v}{1+\omega^2} & 0
\end{pmatrix},
\end{align}
with $\ell_m{}^a$ and $\ell_{ma}$ being the matrix elements of $\ell_m^a$ and $\ell_{am}$\,, the background fields are found to be
\begin{align}
 \rmd s^2 = \tilde{\alpha}'\,k\,\bigl( \rmd u^2 + 2\cos2\rho\,\rmd u\,\rmd v+ \rmd v^2 + \rmd \rho^2\bigr)\,,\qquad
 B_2 = \tilde{\alpha}'\,k \cos2\rho \,\rmd u\wedge\rmd v\,.
\label{eq:g-B-ex1}
\end{align}
We choose the ranges of the coordinates as 
\begin{align}
 0\leq u\leq \pi \,,\qquad 
 0\leq v\leq 2\pi \,,\qquad 
 0\leq \rho\leq \tfrac{\pi}{2}\,.
 \label{eq:su2range}
\end{align}

The left/right currents, which correspond to the ones defined in section \ref{sec:CA-WZW} with $\gamma=-1/\tilde{\alpha}'$, are denoted by
\begin{align}
 \cJ_{\hat{\gga}} = -\frac{k}{2} \,\bigl(\partial_+ f\, f^{-1}\bigr)_{\hat{\gga}} \,,\qquad 
 \bar{\cJ}_{\check{\gga}} = \frac{k}{2} \,\bigl(f^{-1}\, \partial_- f\bigr)_{\check{\gga}}\,,
\label{eq:ex1-J-def}
\end{align}
where we have used the equations of motion $\partial_- f\, f^{-1}=0=f^{-1}\, \partial_+ f$\,.
We have also introduced the generators
\begin{align}
 t_{\check{1}} = - t_{\hat{1}}\,,\qquad 
 t_{\check{2}} = t_{\hat{2}}\,,\qquad 
 t_{\check{3}} = - t_{\hat{3}}\,,
\end{align}
and expanded the right currents by using $t_{\check{\gga}}$\,, so that the left and the right currents take the symmetric form
\begin{align}
\begin{alignedat}{2}
 \cJ_{\hat{3}} &= k\,\bigl(\partial_+ u + \cos 2\rho \,\partial_+ v \bigr)\,,&
 \cJ_{\hat{\pm}} &= k\,\bigl(\partial_+\rho \pm i \sin 2\rho \,\partial_+ v \bigr)\,\Exp{\mp 2iu}\,,
\\
 \bar{\cJ}_{\check{3}} &= k\,\bigl(\partial_- v + \cos 2\rho \,\partial_- u \bigr)\,,\qquad &
 \bar{\cJ}_{\check{\pm}} & = k\,\bigl(\partial_-\rho \pm i \sin 2\rho \,\partial_- u \bigr)\,\Exp{\mp 2iv}\,,
\end{alignedat}
\label{eq:JJbar-ex1}
\end{align}
where $\cJ_{\hat{\pm}} \equiv \cJ_{\hat{1}}\pm i \cJ_{\hat{2}}$ and $\bar{\cJ}_{\check{\pm}}\equiv \bar{\cJ}_{\check{1}}\pm i \bar{\cJ}_{\check{2}}$\,. 
Since $t_{\hat{\gga}} \to t_{\check{\gga}}$ is an automorphism of $\mathfrak{su}(2)$, the Poisson brackets among $\bar{\cJ}_{\check{\gga}}$ are of the same form as those among $\bar{\cJ}_{\hat{\gga}}$ obtained by using $t_{\hat{\gga}}$.

Due to the self-duality constraint \eqref{eq:r-tilde2}, the right currents are expressed as
\begin{align}
 \bar{\cJ}_{\check{\pm}} = \pm i\,k\,\frac{(1+\omega^2)\,(\omega\pm i)^2}{8\,\omega}\,\bigl(\tilde{r}_{1-} \mp i\,\tilde{r}_{2-} \bigr)\,,
\qquad 
 \bar{\cJ}_{\check{3}} = -k\,\frac{1+\omega^2}{4}\,\tilde{r}_{3-}\,.
\end{align}
Since
\begin{align}
 \tilde{r}_1 \pm i\,\tilde{r}_2 = \Exp{\frac{\omega(1\mp i \omega)}{1+\omega^2}\,\tilde{\rho}}\bigl(\rmd \tilde{x} \pm i\,\rmd \tilde{y} \bigr)\,, \qquad
 \tilde{r}_3 = \frac{\rmd \tilde{\rho}}{1+\omega^2}\,,
\end{align}
the above currents also become
\begin{align}
 \bar{\cJ}_{\check{\pm}} = \pm i\,k\,\frac{(1+\omega^2)\,(\omega\pm i)^2}{8\,\omega}\,\Exp{\frac{\omega\,(1\pm i\omega)}{1+\omega^2}\,\tilde{\rho}} \,\bigl(\partial_-\tilde{x}\mp i\,\partial_- \tilde{y}\bigr) \,,
\qquad
 \bar{\cJ}_{\check{3}} = - \frac{k}{4}\,\partial_- \tilde{\rho}\,.
\label{eq:cJ-dual-coords}
\end{align}
We note that $\partial_+ \tilde\rho =0$. 
The energy-momentum tensors in \eqref{eq:TTbar-def} are
\begin{align}
 T_{++} = -\frac{1}{k}\,\delta^{\hat{\gga}\hat{\ggb}}\, \cJ_{\hat{\gga}}\,\cJ_{\hat{\ggb}} \,,\qquad
 T_{--} = -\frac{1}{k}\,\delta^{\check{\gga}\check{\ggb}}\, \bar{\cJ}_{\check{\gga}}\,\bar{\cJ}_{\check{\ggb}} \,.
\end{align}
In the quantum case, $k$ in the denominator is shifted by the second Casimir of $\mathfrak{su}(2)$ as $k+2$, though it is irrelevant classically since $k \gg 1$.

\subsubsection{Dual model}
\label{sec:ex1-dual}

Now, we consider a Poisson--Lie T-duality transformation which exchanges $T_3$ and $T^3$\,. 
Under this transformation, $\hat{\cH}_{AB}$ is not changed, but the structure constants become
\begin{align}
\begin{split}
 f'_{13}{}^{1} &= \omega\,,\quad
 f'_{13}{}^{2} = \omega^2\,,\quad
 f'_{23}{}^{1} = -\omega^2\,,\quad
 f'_{23}{}^{2} = \omega\,,
\\
 f'_{1}{}^{13}&= \omega\,,\quad
 f'_{2}{}^{13}=-1\,,\quad
 f'_{1}{}^{23}= 1\,,\quad
 f'_{2}{}^{23}= \omega\quad (\omega>0)\,.
\end{split}
\end{align}
This corresponds to the Manin triple $(\bm{7_{1/\omega}}\vert \bm{7_\omega}\vert \bm{-\omega^2})$\,. 
Therefore, at least for $\omega\neq 1$, this is a non-trivial example of Poisson--Lie T-duality, which changes the coset $\cD/\tilde{G}_{1/\omega}$ into another coset $\cD/\tilde{G}_\omega$ with $\tilde{G}_\omega$ corresponding to $\bm{7_\omega}$. 

In this case, we define the generators $t'_{\hat{\gga}}\equiv -\frac{1}{1+\omega^2}\,\delta_{\hat{\gga}b}\,\bigl(T'^{b} + T'_c\,E_+^{cb}\bigr)$.
We then find 
\begin{align}
 t'_{\hat{1}} = t_{\hat{1}}\,,\qquad
 t'_{\hat{2}} = t_{\hat{2}}\,,\qquad
 t'_{\hat{3}} = t_{\hat{3}}\,,
\end{align}
which generate the same Lie algebra $\mathfrak{r}$ as above.
It turns out that the following parametrization of $l$ is useful for the discussion below,
\begin{align}
 l = f'\,\tilde{g}'\,, \qquad
 f' \equiv \Exp{-2\,u'\,t_{\hat{3}}}\Exp{-2\,\rho'\,t_{\hat{1}}}\Exp{-2\,v'\,t_{\hat{3}}}\,,\qquad 
 \tilde{g}' \equiv \Exp{\frac{\tilde{\rho}'}{1+\omega^2}\,T_3}\Exp{\tilde{y}'\,T^2}\Exp{\tilde{x}'\,T^1} \,,
 \label{eq:f'tildeg'}
\end{align}
which gives the same background fields \eqref{eq:g-B-ex1} and the currents \eqref{eq:JJbar-ex1} with all the coordinates replaced with the primed ones. 

\subsubsection{Relation}
Given the above parametrization, we can now find an explicit relation between the original and the dual model.
To this end, we first recall the parameterization
\begin{align}
 f\,\tilde{g} = l = f'\,\tilde{g}'\,.
\end{align}
In general, such a dual parametrization gives a highly non-linear relation between the coordinates of the dual pair.
In our example, however, the relation turns out to be very simple; 
\begin{align}
 \tilde{\rho}' = - \tilde{\rho}\,,\qquad
 v'= v + \tfrac{1}{2}\,\tilde{\rho}\,,
\label{eq:coords-relation}
\end{align}
and the other coordinates are invariant. 
Moreover, from $[T_3,\,T^3]=0$, it follows that
\begin{align}
 f' = f \Exp{\frac{\tilde{\rho}}{1+\omega^2}\,T^3-\frac{\tilde{\rho}'}{1+\omega^2}\,T_3} = f\,\Exp{-\tilde{\rho}\,t_{\hat{3}}}\,.
\label{eq:ff'}
\end{align}
The relation between the currents are thus found to be
\begin{align}
\begin{alignedat}{2}
 \cJ'_{\hat{3}} &= \cJ_{\hat{3}} \,,&
 \cJ'_{\hat{\pm}} & = \cJ_{\hat{\pm}} \,,
\\
 \bar{\cJ}'_{\check{3}} &= - \bar{\cJ}_{\check{3}} \,,\qquad &
 \bar{\cJ}'_{\check{\pm}} &= \Exp{\mp i\tilde{\rho}} \bar{\cJ}_{\check{\pm} -} \,.
\end{alignedat}
\label{eq:J-dual}
\end{align}

\subsubsection{Current algebra}
\label{sec:currentalg}

Now we consider the current algebra discussed in section \ref{sec:current-alg} and \ref{sec:CA-WZW}. 
Using Eq.~\eqref{eq:J-alg} and recalling $\gamma=-1/\tilde{\alpha}'$, we find the current algebra in the original model
\begin{align}
\begin{split}
 \{\cJ_{\hat{\gga}}(\sigma^+),\,\cJ_{\hat{\ggb}}(\sigma^{\prime +})\} &= - 2\pi f_{\hat{\gga}\hat{\ggb}}{}^{\hat{\ggc}}\,\cJ_{\hat{\ggc}}(\sigma^+)\,\delta(\sigma^+-\sigma^{\prime +}) + \pi k \, \delta_{\hat{\gga}\hat{\ggb}}\,\delta'(\sigma^+-\sigma^{\prime +})\,,
\\
 \{\bar{\cJ}_{\check{\gga}}(\sigma^-),\,\bar{\cJ}_{\check{\ggb}}(\sigma^{\prime -})\} &= - 2\pi f_{\check{\gga}\check{\ggb}}{}^{\check{\ggc}}\,\bar{\cJ}_{\check{\ggc}}(\sigma^-)\,\delta(\sigma^--\sigma^{\prime -}) + \pi k \,\delta_{\check{\gga}\check{\ggb}}\,\delta'(\sigma^--\sigma^{\prime -})\,,
\end{split}
\label{eq:PBsu2affineLie}
\end{align}
where we have used $\hh_{\hat{\gga}\hat{\ggb}}= \frac{\tilde{\alpha}'\,k}{2}\,\delta_{\hat{\gga}\hat{\ggb}}$\,. 
More explicitly, for the right currents, it is written as 
\begin{align}
\begin{split}
 \{\bar{\cJ}_{\check{3}}(\sigma^-),\,\bar{\cJ}_{\check{3}}(\sigma^{\prime -})\}&= \pi k\,\delta'(\sigma^--\sigma^{\prime -})\,,
\\
 \{\bar{\cJ}_{\check{3}}(\sigma^-),\,\bar{\cJ}_{\check{\pm}}(\sigma^{\prime -})\}&= \pm 2\pi i\, \bar{\cJ}_{\check{\pm}}(\sigma^-)\,\delta(\sigma^--\sigma^{\prime -})\,,
\\
 \{\bar{\cJ}_{\check{+}}(\sigma^-),\,\bar{\cJ}_{\check{-}}(\sigma^{\prime -})\}&= 4 \pi i\,\bar{\cJ}_{\check{3}}(\sigma^-)\,\delta(\sigma^--\sigma^{\prime -}) + 2 \pi k \, \delta'(\sigma^--\sigma^{\prime -})\,.
\end{split}
\label{eq:ex1-current-alg}
\end{align}
These form the $\widehat{\mathfrak{su}}(2)$ affine Lie algebra in the sense of Poisson brackets.

As for the currents of the dual model, since the dual model is again the SU(2) WZNW model, 
the Poisson brackets are also 
of the form in \eqn{eq:PBsu2affineLie} and \eqn{eq:ex1-current-alg} in terms of the brackets defined through the dual action.
Given the relations \eqn{eq:J-dual} and \eqn{eq:cJ-dual-coords}, 
the Poisson brackets of the original model among $\tilde{\cJ}'_{\check{\gga}}$ 
are also computed, up to those involving the zero mode of $\tilde\rho$.
These should be obtained from the sigma model on the Drinfel'd double described by \eqn{eq:SM-action22} with the constraint \eqn{eq:SD} (or the $\cE$-model).
However, there is a simple way around: as mentioned above, Poisson--Lie T-duality is represented by a canonical transformation, which means that the Poisson brackets of the original model among $\tilde{\cJ}'_{\check{\gga}}$ are again of the same form as in \eqn{eq:PBsu2affineLie} and \eqn{eq:ex1-current-alg}.
These in turn determine the brackets involving the zero mode of $\tilde\rho$. 

Let us explicitly check this below. 
First, from \eqref{eq:J-dual} it follows that
\begin{align}
 \{\bar{\cJ}'_{\check{3}}(\sigma^-),\,\bar{\cJ}'_{\check{3}}(\sigma^{\prime -})\}
 = \{\bar{\cJ}_{\check{3}}(\sigma^-),\,\bar{\cJ}_{\check{3}}(\sigma^{\prime -})\} = \pi k\,\delta'(\sigma^--\sigma^{\prime -})\,.
\end{align}
Next, a slightly non-trivial bracket is
\begin{align}
 \{\bar{\cJ}'_{\check{3}}(\sigma^-),\,\bar{\cJ}'_{\check{\pm}}(\sigma^{\prime -})\} 
 &= -\{\bar{\cJ}_{\check{3}}(\sigma^-),\,\Exp{\mp i\tilde{\rho}(\sigma^{\prime -})}\bar{\cJ}_{\check{\pm}}(\sigma^{\prime -})\} 
\nn\\
 &= \mp 2\pi i\, \bar{\cJ}'_{\check{\pm}}(\sigma^-)\,\delta(\sigma^--\sigma^{\prime -})
 \pm i\,\{\bar{\cJ}_{\check{3}}(\sigma^-),\,\tilde{\rho}(\sigma^{\prime -})\}\,\bar{\cJ}'_{\check{\pm}}(\sigma^{\prime -})\,. 
\end{align}
Requiring the bracket to have the same form as Eq.~\eqref{eq:ex1-current-alg}
in accordance with the canonical transformation,
we should have
\begin{align}
 \{\bar{\cJ}_{\check{3}}(\sigma^-),\,\tilde{\rho}(\sigma^{\prime -})\}
 = 4\pi\, \delta(\sigma^--\sigma^{\prime -})\,.
\label{eq:J3-rho}
\end{align}
This is confirmed to be 
consistent with $\{\bar{\cJ}_{\check{3}}(\sigma^-),\,\bar{\cJ}_{\check{3}}(\sigma^{\prime -})\} = \pi k\,\delta'(\sigma^--\sigma^{\prime -})$ by acting 
with $\partial'_-$ and using the relation between $\tilde\rho$ and $\bar{\cJ}_{\check{3}}$ in \eqref{eq:cJ-dual-coords}. 
If we consider a mode expansion
\begin{align}
 \bar{\cJ}_{\check{3}} = \sum_{n\in\mathbb{Z}}\bar{\alpha}_{n}\Exp{-in\sigma^-}\,,
 \label{eq:J3modes}
\end{align}
where $\bar{\alpha}_{n}$ obey the algebra
\begin{align}
 \{\bar{\alpha}_{n},\,\bar{\alpha}_{m}\} = - \frac{ik}{2}\,n\,\delta_{n+m,0}\,,
\end{align}
we can expand $\tilde{\rho}(\sigma^-)$ as
\begin{align}
 \tilde{\rho}(\sigma^-) = \tilde{\rho}_0 - \frac{4}{k}\, \bar{\alpha}_0\,\sigma^-
 - \frac{4}{k} \sum_{n\neq 0}\frac{i}{n}\,\bar{\alpha}_{n}\Exp{-in\sigma^-}\,,
 \label{eq:expandrho}
\end{align}
by introducing the zero mode $\tilde{\rho}_0$\,. 
In terms of the modes, the only non-trivial requirement 
for \eqref{eq:J3-rho} is
\begin{align}
 \{\tilde{\rho}_0,\,\bar{\alpha}_n\} = - 2\,\delta_{0,n}\,.
 \label{eq:tilderho0PB1}
\end{align}
Similarly, we find 
\begin{align}
 \{\bar{\cJ}'_{\check{+}}(\sigma^-),\,\bar{\cJ}'_{\check{-}}(\sigma^{\prime -})\}&= 
 \{\Exp{-i\tilde{\rho}(\sigma^-)}\bar{\cJ}_{\check{+}}(\sigma^-),\,\Exp{i\tilde{\rho}(\sigma^{\prime -})}\bar{\cJ}_{\check{-}}(\sigma^{\prime -})\}
\nn\\
 &= 4 \pi i\,\bar{\cJ}'_{\check{3}}(\sigma^-)\,\delta(\sigma^--\sigma^{\prime -}) 
 + 2 \pi k \, \delta'(\sigma^--\sigma^{\prime -}) 
\nn\\
 &\quad 
 -i \bar{\cJ}'_{\check{+}}(\sigma^-)\,\{ \tilde{\rho}(\sigma^-) ,\,\bar{\cJ}_{\check{-}}(\sigma^{\prime -})\}\,\Exp{i\tilde{\rho}(\sigma^{\prime -})}
\nn\\
 &\quad
  + i \Exp{-i\tilde{\rho}(\sigma^-)}\{\bar{\cJ}_{\check{+}}(\sigma^-),\, \tilde{\rho}(\sigma^{\prime -})\}\,\bar{\cJ}'_{\check{-}}(\sigma^{\prime -})
\nn\\
 &\quad + \bar{\cJ}'_{\check{+}}(\sigma^-)\,\{\tilde{\rho}(\sigma^-) ,\, \tilde{\rho}(\sigma^{\prime -}) \}\,\bar{\cJ}'_{\check{-}}(\sigma^{\prime -})\,.
\end{align}
For the last three lines to be canceled out, we should have 
\begin{align}
 \{\tilde{\rho}_0,\,\tilde{\rho}_0\} = 0\,,\qquad 
 \{\tilde{\rho}_0,\,\bar{\cJ}_{\check{\pm}}(\sigma^-)\} = \pm \frac{4i}{k}\,\sigma^-\,\bar{\cJ}_{\check{\pm}}(\sigma^-)\,. 
 \label{eq:tilderho0PB2}
\end{align}
These give
\begin{align}
 \{\tilde{\rho}(\sigma^-),\,\tilde{\rho}(\sigma^{\prime -})\}
 &= -\frac{16\pi}{k}\,\theta(\sigma^--\sigma^{\prime -})\,,
\label{eq:rho-rho}
\\
 \{\tilde{\rho}(\sigma^-),\,\bar{\cJ}_{\check{\pm}}(\sigma^{\prime -})\}
 &= \mp \frac{8\pi i}{k}\,\theta(\sigma^--\sigma^{\prime -})\,\bar{\cJ}_{\check{\pm}}(\sigma^{\prime -})\,,
\label{eq:rho-Jpm}
\end{align}
where $\theta(\sigma)=\frac{1}{2\pi}\,(\sigma - \sum_{n\neq 0}\frac{i}{n}\,\Exp{in\sigma})$ is the step function.

Consequently, the currents in the dual model indeed satisfy the same algebra even in terms of the Poisson bracket of the original model.
We thus find that the duality transformation contrives to induce an automorphism of the current algebra.
Turning the argument around and assuming the Poisson brackets 
involving the zero mode of $\tilde\rho$ give another derivation of the 
canonical transformation in our case of Poisson--Lie T-duality.
The Poisson bracket \eqref{eq:rho-rho} means that $\tilde\rho(\sigma^-)$ is a chiral free boson, as is understood also from the relation to $\bar{\cJ}_{\check{3}}$ in \eqn{eq:cJ-dual-coords}.
We discuss that implication further in section \ref{sec:quantum}.

Summarizing, by finding appropriate coordinates in the dual model as in \eqref{eq:f'tildeg'}, we have found a simple relation of the currents \eqref{eq:J-dual}.
In terms of the currents, the relation is non-linear and non-local because $\tilde{\rho}$ is expressed by an integral of $\bar{\cJ}_{\check{3}}$ through \eqref{eq:cJ-dual-coords}. 
It also involves an infinite number of the modes of the currents, which are defined as in \eqn{eq:J3modes}.
The zero mode $\tilde\rho_0$ is not given by the currents.
The transformation is however found to be an automorphism of the current algebra, as assured by the general property that Poisson--Lie T-duality is expressed as a canonical transformation. 
Given these results, we can explore the quantum aspects of the duality, including global issues and details of the spectrum, as discussed in section \ref{sec:quantum}.
We note that due to the properties mentioned above the automorphism is 
not on the list of the classification genuinely in terms of 
$\widehat{\mathfrak{su}}(2)_k$ \cite{Kac-Peterson,Moody-Pianzola}; 
to analyze it, the space
that the currents act on should be properly defined, 
or a completion should be necessary. 

\subsection{Example 2}

Before moving on to the quantum analysis, we note that the limiting case $\omega\to 0$ in Example 1, which is singular in Eq.~\eqref{eq:E-ex1}, can be achieved by considering another $2d$-dimensional Lie algebra, where the structure constants contain $f_{abc}$ as well and are given by
\begin{align}
 f_{12}{}^3=-1\,,\quad
 f_{23}{}^1=-1\,,\quad
 f_{31}{}^2=-1\,,\quad
 f_{123}=1\,.
\end{align}
This is generalizing the Drinfel'd double in the sense that $f_{abc} \neq 0$.
Here we choose $E_+^{ab}$ as
\begin{align}
 E_+^{ab} = \begin{pmatrix} 1 & 0 & 0 \\ 0 & 1 & 0 \\ 0 & 0 & 1 \end{pmatrix}.
\end{align}
In this case, the action of the WZNW model becomes
\begin{align}
 S_{\text{\tiny WZNW}}(f)\equiv S = - \frac{k}{8\pi} \int_\Sigma \kappa(\ell, *\ell) + \frac{k}{24\pi}\int_{\cB} \kappa( \ell,\,[\ell,\,\ell]) \,,
\end{align}
where we have defined the level $k\equiv \frac{4}{\tilde{\alpha}'}$ and a bilinear form $\kappa(t_{\hat{\gga}},\,t_{\hat{\ggb}})=-\frac{1}{2}\,\delta_{\hat{\gga}\hat{\ggb}}$\,. 
The generators $t_{\hat{\gga}}\equiv - \delta_{\hat{\gga}b}\,\bigl(T^{b} + T_c\,E_+^{cb}\bigr)$, which span $\mathfrak{r}$\,, satisfy the $\mathfrak{su}(2)$ algebra
\begin{align}
 [t_{\hat{1}},\,t_{\hat{2}}] = t_{\hat{3}}\,,\qquad
 [t_{\hat{2}},\,t_{\hat{3}}] = t_{\hat{1}}\,,\qquad
 [t_{\hat{3}},\,t_{\hat{1}}] = t_{\hat{2}}\,.
\end{align}
Here again we consider a Poisson--Lie T-duality transformation which exchanges $T_3$ and $T^3$\,,
under which the constant part of the generalized metric $\hat{\cH}_{AB}$ is not changed, and
the structure constants become
\begin{align}
 f'_{12}{}^{3}=1\,,\quad
 f'_{1}{}^{23}=-1\,,\quad
 f'_{2}{}^{13}= 1\,,\quad
 f'_{123}=-1\,.
\end{align}
In the original model the maximally isotropic subalgebra $\tilde{\mathfrak{g}}$ is Abelian while in the dual model it is a Lie algebra of Bianchi type $\bm{7_0}$\,, and thus this Poisson--Lie T-duality gives a map between two different cosets. 

In the original/dual model, we can parameterize $l$ as
\begin{alignat}{2}
 l &= f\,\tilde{g} = f'\,\tilde{g}'\,,& 
\\
 f& \equiv \Exp{-2\,u\,t_{\hat{3}}}\Exp{-2\,\rho\,t_{\hat{1}}}\Exp{-2\,v\,t_{\hat{3}}}\,,\qquad&
 \tilde{g}&\equiv \Exp{\tilde{\rho} \,T^3}\Exp{\tilde{y} \,T^2}\Exp{\tilde{x} \,T^1}\,,
\\
 f'& \equiv \Exp{-2\,u\,t_{\hat{3}}}\Exp{-2\,\rho\,t_{\hat{1}}}\Exp{-2\,v'\,t_{\hat{3}}}\,,\qquad&
 \tilde{g}'&\equiv \Exp{\tilde{\rho}' \,T_3}\Exp{\tilde{y} \,T^2}\Exp{\tilde{x} \,T^1}\,,
\end{alignat}
where the new coordinates $\tilde\rho'$ and $v'$ are given by
\begin{align}
 \tilde{\rho}'=-\tilde\rho\,,
 \qquad v'= v + \tfrac{1}{2}\,\tilde{\rho}\,.
 \label{eq:coords-relation2}
\end{align}
Again we define the left/right currents as
\begin{align}
 \cJ_{\hat{\gga}} = - \frac{k}{2} \,\bigl(\partial_+ f\, f^{-1}\bigr)_{\hat{\gga}} \,,\qquad 
 \bar{\cJ}_{\check{\gga}} = \frac{k}{2} \,\bigl(f^{-1}\, \partial_- f\bigr)_{\check{\gga}}\,,
\end{align}
and obtain the same expression as Eq.~\eqref{eq:JJbar-ex1} with $k= \frac{4}{\tilde{\alpha}'}$ (or $\omega=0$). 
Then the relation of the currents between the original and the dual model becomes
\begin{align}
\begin{alignedat}{2}
 \cJ'_{\hat{3}} &= \cJ_{\hat{3}} \,,&
 \cJ'_{\hat{\pm}} & = \cJ_{\bar{\pm}} \,,
\\
 \bar{\cJ}'_{\check{3}} &= - \bar{\cJ}_{\check{3}} \,,\qquad &
 \bar{\cJ}'_{\check{\pm}} & = \Exp{\mp i\tilde{\rho}} \bar{\cJ}_{\check{\pm}} \,,
\end{alignedat}
\label{eq:J-dual2}
\end{align}
which is precisely the same as the one in the previous example 
in \eqref{eq:J-dual}.
Accordingly, the current algebra is analyzed as in Example 1, and we find that the duality transformation induces a non-local chiral automorphism.

\subsection{Action on geodesic solutions}

Usual Abelian T-duality exchanges momentum and winding modes.
Similarly, our Poisson--Lie T-duality can exchange the roles of the world-sheet coordinates $\tau$ and $\sigma$, though it does not generate true windings because SU(2) is simply connected.
To see this, let us consider the action of the duality on the zero mode represented by geodesic solutions. 

Since the general solution of the WZNW model takes the form
$f = f_+(\sigma^+) f_-(\sigma^-)$, the geodesics represented, for example, 
by $f=g_{\hat{\gga}}(p\,\tau)$ with $p$ being constant provide 
solutions, 
where $g_{\hat{\gga}}(\tau) \equiv \exp[ \tau\,t_{\hat{\gga}}] 
= \exp[ \sigma^+\, t_{\hat{\gga}}/2] \exp[ \sigma^-\, t_{\hat{\gga}}/2]$. 
For $f=g_{\hat{\ggb}}(p\, \tau)$, the currents and the energy-momentum tensor in the right mover are
\begin{align}
 \bar{\cJ}_{\check{\gga}} =(-1)^{\check{\gga}}\, \frac{k}{4}\, p \, \delta_{\check{\gga}\hat{\ggb}} \comma \qquad
 T_{--} = - \frac{k}{16}\, p^2 \period
\end{align}
As a solution in string theory, the value of $p$ has to satisfy the Virasoro constraints including other part of spacetime. 
In particular, taking $f=g_{\hat{3}}(p\,\tau)$, one has $\tilde{\rho} = \tilde{\rho}_0 + p\,\sigma^-$ from \eqn{eq:cJ-dual-coords}.
The duality transformation \eqn{eq:ff'} and the corresponding one in Example 2 (where $\omega=0$ should be understood) are thus
\begin{align}
 f' = g_{\hat{3}}(p\,\tau) \Exp{-p\,\sigma^- t_{\hat{3}}} \Exp{-\tilde{\rho}_0\, t_{\hat{3}}}
 \sim g_{\hat{3}}(p\,\sigma) \comma
\end{align}
up to the zero mode part associated with $\tilde{\rho}_0$ that can be absorbed into a shift of $\sigma$\,. 
This indeed shows the exchange of the roles of $\tau$ and $\sigma$. 
As discussed in section \ref{sec:global}, the value of $p$ is found to be quantized in the unit of $2/k$ in the quantum case, which specifies the global structure of the dual target space.

The duality transformation is trivial for $f=g_{\hat{\gga}}$ $(\hat{\gga} =1,2)$, since $\bar{\cJ}_{\check{3}} = 0$ in these cases. 
One can also consider the general geodesic $f= U g_{\hat{3}}(p\, \tau) V^{-1}$ where $U$ and $V$ are constant elements of SU(2). 
In this case, the $t_{\hat{3}}$-component of $V t_{\hat{3}} V^{-1}$ contributes to $\bar{\cJ}_{\check{3}}$.

\section{Quantum description of duality}
\label{sec:quantum}
In the previous section, we found concrete examples where the SU(2) WZNW model is self-dual under Poisson--Lie T-duality. 
The duality was expressed as a chiral automorphism 
of the current algebra or the $\widehat{\mathfrak{su}}(2)$ affine Lie algebra.

In this section, we show that this automorphism is extended to the quantum or the $\alpha'$-exact one, 
which induces an isomorphism of the CFT associated with the SU(2) WZNW model.
We thus find that the classical duality in the previous section 
is promoted to the quantum duality which ensures the full quantum equivalence of the dual pair. 
In the course, the global structure of the duality is also figured out.
From the world-sheet point of view, a quantum equivalence under any duality may be reduced to an isomorphism 
of the underlying CFT, which also implies an automorphism of the symmetry algebras, as for usual Abelian T-duality.
Our case of Poisson--Lie T-duality is regarded as providing another example.
In the following, we mainly focus on Example 1, but the discussion also applies to Example 2.

\subsection{Parafermionic formulation of $\widehat{\mathfrak{su}}(2)$ affine Lie algebra}
For notational convenience, in what follows we take the world-sheet to be Euclidean by performing the Wick rotation $\tau=-it$, and define the complex coordinates 
\begin{align}
z \equiv \Exp{t+i\sigma} = \Exp{i \sigma^+}\,,\qquad
 \bar{z} \equiv \Exp{t-i\sigma }= \Exp{i \sigma^-}\,.
 \label{eq:zzbar}
\end{align}
We also introduce the currents $J_{\hat{\mathfrak{a}}}(z)$ and $\bar{J}_{\check{\mathfrak{a}}}(\bar{z})$ in the Hermitian basis 
$\mathfrak{t}_{\hat{\mathfrak{a}}}\equiv i\,t_{\hat{a}}$ and 
$\mathfrak{t}_{\check{\mathfrak{a}}}\equiv i\,t_{\check{a}}$; 
\begin{align}
J_{\hat{\mathfrak{a}}}(z) = -i \cJ_{\hat{a}z}(z) 
= -i\frac{\rmd\sigma^+}{\rmd z}
\cJ_{\hat{a}+}(\sigma^+)\,, \qquad
\bar{J}_{{\check{a}}}(\bar{z}) = -i \bar{\cJ}_{\check{a}\bar{z}}(\bar{z}) 
= -i \frac{\rmd\sigma^-}{\rmd\bar{z}}
\bar{\cJ}_{\check{a}-}(\sigma^-)\,.
 \label{eq:JahatJa}
\end{align}
Replacing the Poisson brackets of the modes of the currents with the commutators as $i \, \{ \cdot , \cdot\} \to [\cdot, \cdot]$, the current algebra in \eqn{eq:ex1-current-alg} is translated to the operator product expansions (OPEs) for the affine Lie algebra $\widehat{\mathfrak{su}}(2)_k$,
\begin{align}
  J_3(z) J_3(w) &\sim \frac{k/2}{(z-w)^2}
  \comma \qquad
  J_3(z) J_\pm(w) \sim \frac{\pm J_\pm(w)}{z-w}
  \comma \nn \\
  J_+(z) J_-(w) & \sim \frac{k}{(z-w)^2} + \frac{2 J_3(w)}{z-w}
  \period
  \label{eq:su2AffineLie}
\end{align}
Here we have set $ J_3(z) \equiv J_{\hat{\mathfrak{3}}}(z) $,
$J_\pm(z) \equiv J_{\hat{\mathfrak{1}}}(z) \pm i J_{\hat{\mathfrak{2}}}(z) $.
We have similar OPEs for $\bar{J}_3(\zbar) = \bar{J}_{\check{\mathfrak{3}}}(\bar{z})$,
$\bar{J}_\pm(\zbar) = \bar{J}_{\check{\mathfrak{1}}}(\bar{z}) \pm i \bar{J}_{\check{\mathfrak{2}}}(\bar{z})$. 

The automorphism of $\widehat{\mathfrak{su}}(2)_k$ found in the previous section 
turns out to have a simple representation, though it might appear to be intricate, 
as suggested by the fact that $\tilde\rho$ in \eqn{eq:cJ-dual-coords} is a chiral free boson.
To see this, let us introduce the parafermionic formulation of the affine Lie algebra \cite{Fateev:1985mm,Gepner:1986hr}.
At level $k$, the affine Lie algebra $\widehat{\mathfrak{su}}(2)_k$ 
in the holomorphic sector is represented by a free boson $\varphi(z)$
and the parafermions $\psi_1(z), \psi_1^\dagger(z)$ as
\begin{align}
\label{eq:su2pf}
  J_3(z) &= i \sqrt{k \over 2} \, \del \varphi(z) \comma \nn \\
  J_+(z) &= \sqrt{k}\,\psi_1(z) \Exp{i \sqrt{2 \over k} \, \varphi(z)} \comma \quad
  J_-(z) = \sqrt{k}\,\psi_1^\dagger(z) \Exp{-i \sqrt{2 \over k} \, \varphi(z)} \period
\end{align}
One can check that the OPEs \eqn{eq:su2AffineLie} are indeed recovered by using 
\begin{align}
\label{eq:pfOPE}
 \varphi(z) \varphi(w) &\sim - \ln(z-w) \comma \nn \\
 \psi_1(z)\psi_1^\dagger(w) &\sim (z-w)^{\frac{2}{k}-2} \comma \quad
 \psi_1^\dagger(z)\psi_1(w) \sim (z-w)^{\frac{2}{k}-2} \period
\end{align}
For $k=1$, the parafermions are absent. 
For $k=2$, they reduce to free fermions. 
We have similar expressions for the anti-holomorphic sector, which are denoted by bars, 
for example, as $\bar{\varphi}(\zbar)$.

\subsection{Poisson--Lie T-duality as isomorphism of CFT}

From  $\bar{\cJ}_{\check{3}}$ in \eqn{eq:cJ-dual-coords} and 
$\bar{J}_3(\zbar)$ defined as in \eqn{eq:su2pf}, 
one finds that $\tilde\rho(\sigma^-)$ corresponds to $\bar\varphi(\zbar)$. 
Indeed, both $\tilde\rho$ and $\bar\varphi$ are chiral free bosons, 
and the Poisson brackets involving $\tilde\rho$ and its modes
in section \ref{sec:currentalg} are 
in accordance with the OPE of $\bar\varphi(\zbar)$ as in \eqn{eq:pfOPE}.
One can then identify $k\,\tilde\rho/4$ with 
$\sqrt{k/2} \, \bar\varphi$, and hence
the sign-change of $\bar{J}_3(\zbar)$ from \eqn{eq:J-dual} 
with that of $\bar\varphi(\zbar)$.
The transformations of $\bar{\cJ}'_{\check{\pm}}$ in \eqn{eq:J-dual}, 
which preserve the Poisson
brackets of the currents, amount to the change in the sign of $\bar\varphi(\zbar)$
in the parafermionic representation of $\bar{J}_\pm(\zbar)$ defined 
as in \eqn{eq:su2pf}.
The non-locality of the transformations in terms of the currents
has been absorbed in the definition of $\bar\varphi$.

Therefore, the classical Poisson--Lie T-duality transformation 
in \eqn{eq:J-dual} is translated to
\begin{align}
\label{eq:qPLtrans}
  \bar{J}'_3(\zbar) &= - \bar{J}_3(\zbar) 
  =  - i \sqrt{k \over 2} \, \delbar \bar\varphi(\zbar)\comma \nn \\
  \bar{J}'_+(z) &= \Exp{- 2i \sqrt{2 \over k} \, \bar\varphi(\zbar) } \bar{J}_+(\zbar) 
  =  \sqrt{k}\,\bar\psi(\zbar)  \Exp{ -i \sqrt{2\over k} 
  \, \bar\varphi(\zbar)} \comma \\
   \bar{J}'_-(\zbar) &= \Exp{2i \sqrt{2 \over k} \, \bar\varphi(\zbar) } \bar{J}^-(\zbar) 
  =  \sqrt{k}\,\bar\psi^\dagger(\zbar)  \Exp{ i \sqrt{2 \over k} \, \bar\varphi(\zbar)}
  \comma \nn
\end{align}
with the holomorphic currents $J_a(z)$ kept intact.
Compared with the expression of $\bar{J}_a$, the above transformation 
is summarized as the chiral sign-change in all of $\bar{J}_a$,
\begin{align}
 \bar\varphi(\zbar) \to - \bar\varphi(\zbar) \comma
\label{eq:signflip}
\end{align}
where the OPE of $\bar\varphi$ as in \eqn{eq:pfOPE} is kept invariant.
Clearly, this is an automorphism of the parafermionic representation of the SU(2) WZNW model.
This means that the classical Poisson--Lie T-duality
in our case can be promoted to a quantum duality where 
the (self-)dual pair is equivalent in the quantum sense.
As is discussed further below, the spectrum is equivalent,
and so are the correlations functions at any genus of the world-sheet.

We note that the chiral change in the sign, though simple, is not trivial:
In the case of $k=1$, where the parafermions are absent, 
\eqn{eq:signflip} corresponds to 
usual Abelian T-duality of 
a free boson at the self-dual radius which describes the SU(2) WZNW model.
In addition, we recall that the unitary discrete series of the $N=2$ superconformal
field theory (minimal models) are represented by the parafermions
and a free boson \cite{Zamolodchikov:1986gh}. 
The mirror symmetry in the Gepner model consisting of the $N=2$ minimal models 
is implemented also by the chiral change in the sign of the free boson as in \eqn{eq:signflip}
\cite{Greene:1990ud,Greene:1996cy}.\footnote{
The mirror symmetry has been discussed in relation to Poisson--Lie T-duality in \cite{Parkhomenko:1998su}.
}
Furthermore, there exits a marginal deformation 
connecting the SU(2) WZNW model and 
the $N=2$ minimal model, where the radius of the free boson is changed
\cite{Yang:1988bi}. 
Taking this into account, one may say that our Poisson--Lie T-duality
represented by \eqn{eq:signflip} is 
of the same class as the mirror symmetry in the Gepner model.

As in the classical case mentioned above, 
the automorphism of $\widehat{\mathfrak{su}}(2)_k$ in \eqn{eq:qPLtrans} is not generally 
expressed genuinely in terms of the affine Lie algebra.
Thus, the automorphism in our case is not reduced to 
the usual inner or outer automorphism of $\widehat{\mathfrak{su}}(2)_k$ generally.
We discuss related issues later in section \ref{sec:relAbelianT}.

\subsection{Spectrum}
\label{sec:spectrum}

Now, let us see the correspondence of the spectrum in more detail.
To this end, we recall that the primary fields $G_{m,\bar{m}}^{l,\bar{l}}$ 
of the SU(2) WZNW model
at level $k$ are given by the product of those of the parafermionic part
$\Phi_{m,\bar{m}}^{l,\bar{l}}$ and the free-boson part 
\cite{Fateev:1985mm,Gepner:1986hr}
as
\be
  G_{m,\bar{m}}^{l,\bar{l}}(z,\zbar) = \Phi_{m,\bar{m}}^{l,\bar{l}}(z,\zbar) \, 
  \Exp{\frac{i}{\sqrt{2k}} [ m\varphi(z) +\bar{m}\bar\varphi(\zbar) ]}
    \period
  \label{eq:GPhi}
\ee
Here, 
$l = 0,1, ..., k$ and $-l \leq m \leq l$ are respectively 
twice the $\mathfrak{su}(2)$ spin and its component 
in the $3$-direction of the zero-mode algebra of 
$J_{a;0} = \oint {\rmd z \over 2\pi i} J_a(z)$.
Similarly, $\bar{l} $ and $\bar{m} $
are those in the anti-holomorphic sector. 
The general states are obtained by acting with the currents 
in \eqn{eq:su2pf} and those in the anti-holomorphic sector.

Accordingly, the partition functions of the SU(2) WZNW model split into 
the parafermionic (PF) part and the free-boson ($\varphi$) part. For example, 
in the holomorphic sector with fixed $l$ and $m$, one has
\be
\label{Zdecomp}
 Z_{\rm WZNW}^{l,m}(\tau) = Z_{\rm PF}^{l,m}(\tau) Z_\varphi^m(\tau) \comma
\ee
with $\tau$ being the modulus of the world-sheet torus.
The free-boson part is given by
\be
\label{Zvarphi}
 Z_\varphi^m(\tau) = q^{m^2 \over 4k}\eta(\tau)^{-1} \comma
\ee
where $q=\Exp{2\pi i\tau}$ and $\eta(\tau) = q^{1 \over 24} \prod_{n=1}^\infty (1-q^n)$. 
These are customarily expressed by the string functions $c_m^l(\tau)$ as
\be
 Z_{\rm WZNW}^{l,m}(\tau) = q^{m^2 \over 4k } c_m^l(\tau) \comma \qquad
 Z_{\rm PF}^{l,m}(\tau) = \eta(\tau) c_m^l(\tau) \period
\ee
They satisfy
\be
\label{propcml}
  c_m^l = 0 \ \ (l-m \notin 2 \bbZ) \comma \qquad
  c_m^l = c_{-m}^l= c_{m+2k}^l= c_{k+m}^{k-l} \period
\ee
Summing over $m$ and using the periodicity of $c_m^l$ in $m$,
one obtains the character of the spin-$l \over 2$ representation,
\be
  \chi_l (\tau) = \sum_{m=-k+1}^k c_m^l(\tau) \Theta_{m,k}(\tau) \comma
\ee
where $\Theta_{m,k}(\tau) = \sum_{n \in \bbZ} q^{k(n+ \frac{m}{2k})^2}$.
The partition functions in the anti-holomorphic sector are similar.
The total partition functions of the SU(2) WZNW model are then given by 
the modular invariant combinations of the characters of the form,
\be
  Z_{\rm WZNW}(\tau) 
  = \sum_{l,\bar{l}=0}^k N_{l\bar{l}} \, \chi_l(\tau) \chi_{\bar{l}}(\tau)^*
  \comma
\ee
where $N_{l\bar{l}}$ are non-negative integers.
A simple example is the diagonal invariant with
$N_{l\bar{l}} = \delta_{l\bar{l}}$.

Under \eqn{eq:signflip}, the primary fields transform as
\be
  G_{m,\bar{m}}^{l,\bar{l}} 
   \to \Phi_{m,\bar{m}}^{l,\bar{l}} \Exp{\frac{i}{\sqrt{2k}} (m\varphi -\bar{m}\bar\varphi )}
  = \Phi_{m,-\bar{m}'}^{l,\bar{l}} \Exp{\frac{i}{\sqrt{2k}} ( m\varphi +\bar{m}'\bar\varphi )} 
  \label{eq:Gtrans}
 \ee
with $\bar{m}'\equiv -\bar{m}$. 
Together with \eqn{eq:qPLtrans}, the transformations of 
the descendant fields are read off. 
Consequently, 
only the change in the partition function is regarded as 
\be
\label{eq:Zvarphiflip}
Z_{\bar\varphi}^{\bar{m}}(\tau)^* \to Z_{\bar\varphi}^{-\bar{m}}(\tau)^*
\comma
\ee
in the free-boson part of the anti-holomorphic sector, 
which is indeed invariant. 

Alternatively, one may think that, as in the last expression
of \eqn{eq:Gtrans}, the sign of $\bar{m}'$ ($\bar{\bbZ}_{2k}$ charge)
in the parafermionic part is reversed in the total spectrum 
(after $G_{m,\bar{m}}^{l,\bar{l}} \to G_{m,-\bar{m}}^{l,\bar{l}}$),
under which the string function in the anti-holomorphic sector becomes
\be
\label{eq:clmflip}
 c_{\bar{m}}^{\bar{l}}(\tau)^* \to c_{-\bar{m}}^{\bar{l}}(\tau)^* 
 \comma
\ee
which is invariant again.
This corresponds to the order-disorder duality of the 
parafermionic theory \cite{Fateev:1985mm}, 
where the spin and the dual spin fields are exchanged, as is
recognized also in the context of the mirror symmetry 
\cite{Greene:1990ud,Greene:1996cy}.\footnote{
The order-disorder duality has been discussed also in the context of T-duality for example in \cite{Kiritsis:1993ju,Alvarez:1993qi}.
}
Although the spectrum 
is invariant under \eqn{eq:Zvarphiflip} or \eqn{eq:clmflip},
the primary fields of the dual theory 
\eqn{eq:Gtrans} are not of the form \eqn{eq:GPhi}.
This means that the states in the dual theory are generally not expressed 
by those in the original theory, as in Abelian T-duality
connecting different radii of bosons.
This is also in accordance with the fact that the transformations
\eqn{eq:J-dual}, \eqn{eq:J-dual2} or their quantum versions
involve infinitely many modes of the currents, as mentioned above,
and that the order-disorder duality relates equivalent descriptions
which are however non-local to each other.

\subsection{Global structure}
\label{sec:global}
It is important to figure out global properties under dualities in understanding the equivalence of the dual pair.
However, this is generally an open problem in Poisson--Lie T-duality.
Our examples enable us also to address this issue.

From \eqn{eq:expandrho} and the quantization of $\bar{\alpha}_0 \in \bbZ/2$, one finds that $\tilde\rho$ has the periodicity
\begin{align}
 \tilde\rho(\sigma^- +2\pi) = \tilde\rho(\sigma^- ) - \frac{4\pi}{k}\,.
 \label{eq:rhoperiod}
\end{align}
In order for the string in the dual model to be closed, it then follows from \eqref{eq:coords-relation} that the range of the coordinate $v'$ in the dual model should be
\begin{align}
 0\leq v'\leq \frac{2\pi}{k}\,.
\end{align}
This gives $1/k$ times the periodicity which ensures that the dual field $f'$ in \eqn{eq:ff'} is an element of SU(2).
Thus, $f'$ belongs to the lens space SU(2)$/\bbZ_k$ and, precisely speaking, the dual theory is the $\bbZ_k$ orbifold of the level $k$ SU(2) WZNW model.
This picture is consistent with the discussion on the spectrum in section \ref{sec:spectrum}: It is known that the $\bbZ_k$ orbifold
of the $\bbZ_k$ symmetric parafermionic CFT is represented by the order-disorder duality, and that of the level $k$ SU(2) WZNW model is equivalent to the original one
\cite{Gepner:1986hr,Greene:1990ud,Greene:1996cy,Gaberdiel:1995mx,Maldacena:2001ky}. 
In other words, while our Poisson--Lie T-duality is understood as
the chiral change in the sign of $\bar\varphi$ or the oder-disorder
duality from the point of view of the map of the currents, 
it is also understood as the $\bbZ_k$ orbifolding from the point of view of the dual Lagrangian/action.
Requiring the strings be closed also in the total target space of the Drinfel'd double,
one may need to consider them as periodic in $\tilde\rho$ with the given periodicity.
Instead, one may consider, following \cite{Klimcik:2000sk}, ``monodromic strings" which are not necessarily closed in the total target space. 

The isomorphism of the CFT shows that the duality indeed holds in both directions. 
We may also repeat the discussion of the sigma model starting from the dual side, i.e.~the lens space SU(2)$/\bbZ_k$.
For $k=2$, we have SU(2)$/\bbZ_2=$ SO(3).
For $k >2$, the (untwisted) $\widehat{\mathfrak{su}}(2)$ symmetry in the right mover is
broken to $\widehat{\mathfrak{u}}(1)$ \cite{Maldacena:2001ky}.
For example, the right currents 
$\bar{\cJ}_{\check{\pm}}$ in 
\eqn{eq:JJbar-ex1} are not well-defined due to the change of the periodicity of $v$. 
In any case, the orbifold model is equivalent to the one on SU(2), as mentioned above,
after taking into account the twisted sectors. 
The remaining $\widehat{\mathfrak{u}}(1)$ current 
is represented by a free boson whose radius is $1/k$ times the original one.
The quantization of the zero mode of $\tilde\rho$ is also changed.
These are consistent with the duality which maps the dual model back to the original one, though the classical action by itself is not sufficient 
to understand the inverse map from the sigma model point of view because of the twisted sectors.

\subsection{Relation to Abelian T-duality of WZNW models}
\label{sec:relAbelianT}

Our quantum Poisson--Lie T-duality may be regarded as the second simplest to those reducing to usual Abelian T-duality in torus compactifications. 
The duality transformation \eqn{eq:signflip} is the usual T-duality transformation for a free boson representing $\widehat{\mathfrak{u}}(1)$ currents.
These suggest that there may be a simpler formulation.
Indeed, it turns out that the classical transformation \eqn{eq:ff'} can be realized also as Abelian T-duality of the SU(2) WZNW model \cite{Kiritsis:1993ju,Alvarez:1993qi,Alvarez:1994wj}.
Here, we return to the Lorentzian world-sheet to conform to the discussion on the sigma models. 

We begin with the SU(2) WZNW model given by 
\begin{align}
 S_{\text{\tiny WZNW}}(f) = - \frac{k}{8\pi} \int_\Sigma \kappa(\ell, *\ell) + \frac{k}{24\pi}\int_{\cB} \kappa( \ell,\,[\ell,\,\ell]) \,,
\end{align}
where $f \in$ SU(2) and $\ell = f^{-1}\rmd f = \ell^{\hat{\gga}}\, t_{\hat{\gga}}$. 
The notation here is as in section \ref{sec:PLWZW}.
We then gauge an isometry associated with the right-multiplication of $t_{\hat{3}}$ by introducing the gauge field $A\equiv A_+\,\rmd \sigma^+ + A_-\,\rmd \sigma^-$ and an auxiliary field $\tilde{\chi}$ as
\begin{align}
 S_{\text{gauged}} = S_{\text{\tiny WZNW}}(f) - \frac{k}{8\pi}\int_\Sigma\,\rmd\sigma^+\wedge\rmd\sigma^-\, \bigl[A_+ \,A_- + A_+\,(2\, \ell_-^3 - \partial_-\tilde{\chi}) + \partial_+\tilde{\chi}\,A_- \bigr]\,.
\label{eq:gauged-action}
\end{align}
Under a (finite) gauge transformation
\begin{align}
 f \to f\Exp{-\epsilon\,t_{\hat{3}}}\,,\qquad 
 A\to A + \rmd \epsilon\,,\qquad 
 \tilde{\chi}\to \tilde{\chi} - \epsilon\,,
\label{eq:gauge-trnsf}
\end{align}
this action is indeed transformed as 
\begin{align}
 S_{\text{gauged}} \to S_{\text{gauged}} + \frac{k}{8\pi}\int_\Sigma\,\rmd \epsilon \wedge \rmd \tilde{\chi} \,.
\label{eq:transSgauged}
\end{align}
The additional term is a surface term irrelevant to the equations of motion, and it is not important at the classical level.

The equations of motion for $\tilde{\chi}$ in the gauged system \eqn{eq:gauged-action} show that $A$ is closed, and we can choose the gauge $A=0$\, locally.
This reproduces the original model with the action $S = S_{\text{\tiny WZNW}}(f)$\,. 
If we instead use the equations of motion for $A$ and $\hat{A}$\,, we obtain
\begin{align}
 A_+ = -\partial_+\tilde{\chi}\,,\qquad 
 A_- = \partial_-\tilde{\chi} - 2\, \ell_-^3\,,
\label{eq:A-eom}
\end{align}
and the dual action becomes
\begin{align}
 S = S_{\text{\tiny WZNW}}(f) - \frac{k}{8\pi}\int_\Sigma\,\rmd\sigma^+\wedge\rmd\sigma^-\,
\bigl( \partial_+\tilde{\chi}\,\partial_-\tilde{\chi} - 2\,\partial_+\tilde{\chi}\, \ell_-^3 \bigr)\,.
\end{align}
This can be expressed due to the Polyakov--Wiegmann formula as
\begin{align}
 S = S_{\text{\tiny WZNW}}(f')\,,\qquad f'\equiv f\Exp{-\tilde{\chi}\,t_{\hat{3}}} = \Exp{-2\,u\,t_{\hat{3}}}\Exp{-2\,\rho\,t_{\hat{1}}}\Exp{-2\,v'\,t_{\hat{3}}}\,,\qquad
 v' \equiv v+\frac{\tilde{\chi}}{2} \,,
 \label{eq:dualSf'}
\end{align}
where we have again used the parameterization $f=\Exp{-2\,u\,t_{\hat{3}}}\Exp{-2\,\rho\,t_{\hat{1}}}\Exp{-2\,v\,t_{\hat{3}}}$\,.
The dual model is thus described by the action of the SU(2) WZNW model, which
ensures the equivalence of the classical equations of motion.

This duality is promoted to the quantum one by considering the path integral of the gauged system. 
The additional term in \eqn{eq:transSgauged} then has to be a multiple of $2\pi$, not to contribute to the path integral.
This requirement determines the periods of $\tilde\chi$ along the non-trivial homology cycles of $\Sigma$ \cite{Alvarez:1993qi,Gaberdiel:1995mx}. 
In our case, the finite gauge transformation \eqref{eq:gauge-trnsf} corresponds to a coordinate shift $v\to v+\frac{\epsilon}{2}$ and, taking into account 
the periodicity of $v$, the period of $\epsilon$ should be $4\pi$\,.
Then the period of $\tilde{\chi}$ is found as $4\pi/k$, 
by taking $\Sigma$ to be a torus where $\int_\Sigma\,\rmd \epsilon \wedge \rmd \tilde{\chi}=\oint_a \rmd \epsilon \oint_b \rmd \tilde\chi - \oint_a\rmd \tilde\chi \oint_b \rmd \epsilon$ in terms of the cycles $a,b$.

Alternatively, one can fix the gauge where $v=0$. 
Integrating out $\tilde\chi$, one finds that $A$ is pure gauge, $A = \rmd\xi$.
By the summation over the winding numbers, 
the periodicity of $\tilde\chi$ assures that 
the holonomies of $A$ along the cycles are trivial, 
and $A$ is pure gauge globally \cite{Rocek:1991ps,Alvarez:1993qi}.
The resultant action is then 
$S = S_{\text{\tiny WZNW}}(f\Exp{\xi \, t_{\hat{3}}})$.
This shows
that $\xi$ is identified with a coordinate of the original model,
whereas $\tilde\chi$ is identified with one 
of the dual model after integrating out $A$ \cite{Rocek:1991ps}.
Consequently, the additional term in \eqn{eq:transSgauged} is
thought of as 
symmetric with respect to the original and dual models.
The periodicity of $\epsilon$ is also read off from that of $\xi$,
since $\xi$ is regarded as a gauge parameter, $\xi \sim \epsilon$.

Now, we find that the transformation of $f$ in \eqn{eq:dualSf'} coincides with 
that in \eqn{eq:ff'} from our Poisson--Lie T-duality
under the identification $\tilde{\chi} = \tilde{\rho}$.
The relation between the current and $\tilde\rho$ in \eqn{eq:cJ-dual-coords}
is also read off by adopting the gauge $A_\pm = 0$ in \eqn{eq:A-eom}. 
The periodicity of $\tilde\chi$ in the above agrees with 
that of $\tilde\rho$ determined by the quantization 
of $\bar{\alpha}_0$ in section \ref{sec:global}.
This periodicity means that $f'$ is an element of the lens space SU(2)$/\bbZ_k$ as in our case of Poisson--Lie T-duality. 
The quantum equivalence of the dual pair is thus shown similarly as 
in sections \ref{sec:spectrum} and \ref{sec:global}, based on either
on the chiral sing-change/the order-disorder duality, or the $\bbZ_k$ orbifolding.
Indeed, the equivalence of the $\bbZ_k$ orbifold to the original model
has been shown in this context by using the twisted $\widehat{\mathfrak{su}}(2)$ and 
its isomorphism to the untwisted one \cite{Gaberdiel:1995mx}.
Because of the symmetry of the additional term in \eqn{eq:transSgauged} mentioned above, 
the action of the original model on SU(2) with the correct periodicity of the coordinates is obtained 
by the inverse duality transformation from 
that of the dual model on SU(2)$/\bbZ_k$.

We note that generally the duality here is not represented either by the Weyl group of $\widehat{\mathfrak{su}}(2)$, which is implemented as inner automorphisms,
or by other automorphisms genuinely in terms of $\widehat{\mathfrak{su}}(2)$, as argued in the earlier literature.
In this regard, we refer also to \cite{Gaberdiel:1995mx}.
Following the discussion so far, one has to take into account the non-trivial transformation of the right currents as $\bar{\cJ}_{\check{\pm}}$,
and to deal with its non-locality and the issue of the zero mode which is not expressed by the currents of $\widehat{\mathfrak{su}}(2)$ as $\tilde{\rho}_0$.
Both in Poisson--Lie T-duality in section \ref{sec:PLWZW} and in Abelian T-duality of the WZNW model in this subsection, 
these issues are resolved in the parafermionic formulation, or bypassed by considering the orbifold based on the dual Lagrangian.
Therefore, slightly larger frameworks than the untwisted affine Lie algebra $\widehat{\mathfrak{su}}(2)$ itself are necessary to understand the dualities in any case.

For $k=1$, corresponding to the free boson at the self-dual radius, the parafermions
are absent and, for $k=2$, the parafermion reduces
to the usual fermion satisfying $\bar\psi^\dagger_1 = \bar\psi_1$.
In these cases, the transformation of the currents becomes 
$(\bar{J}_3, \bar{J}_\pm) \to (-\bar{J}_3, \bar{J}_\mp)$ in the parafermionic representation, 
which is essentially the Weyl automorphism of $\mathfrak{su}(2)$ up to the signs of $\bar{J}_\mp$.
The Weyl group has been discussed in relation to Abelian T-duality for torus compactifications in \cite{Giveon:1994fu}.

\section{Conclusions}
\label{sec:conclusions}
We have studied Poisson--Lie T-duality of WZNW models. 
When the $2d$-dimensional Lie group $\cD$ contains a $d$-dimensional subgroup $R$ satisfying certain mild conditions, the resultant Poisson--Lie symmetric models become a pair of WZNW models associated with $R$, which means that they are classically self-dual under the duality. 
We have described an explicit construction of the associated currents. 
For this self-duality, we have noted that the conformal invariance can be maintained even when the trace part of the structure constants does not vanish.
As concrete examples, we have considered the self-duality of the SU(2) WZNW model by starting from a six-dimensional Drinfel'd double and its generalization.
Although the transformation of the currents is non-local, non-linear and involving an infinite number of the modes, 
it has been represented as a chiral automorphism of the $\widehat{\mathfrak{su}}(2)$ affine Lie algebra in terms of the Poisson bracket. 

We have found that this classical automorphism can be promoted to the quantum one through the parafermionic
formulation of $\widehat{\mathfrak{su}}(2)$, which induces a non-trivial isomorphism of the WZNW model.
As in usual Abelian T-duality of torus compactifications or in the mirror symmetry of the Gepner model, this isomorphism is represented by the change in the sign of the associated chiral boson or the order-disorder duality of the parafermionic CFT. 
The global structure of the duality has also been figured out.
The duality can be understood also as the one between the SU(2) WZNW model with level $k$ and its $\bbZ_k$ orbifold, which are indeed isomorphic to each other.
Our results provide explicit examples of the Poisson--Lie T-duality transformation which can be promoted to a quantum one and assures the full quantum equivalence of the dual pair to all orders in $\alpha'$ and at any genus, where the issues, for example, of the global structure and the spectrum are also well understood.
These may be regarded as the second simplest to those reducing to usual Abelian T-duality.

Once it is understood that Poisson--Lie T-duality can be further extended to the quantum one, we may expect extensions of our analysis to more general cases.
The relation to Abelian T-duality of WZNW models,
in addition to the classical canonical transformation preserving the underlying affine Lie algebras,  
suggests that the generalization to higher-rank cases may be possible. 
In such cases, the generalized parafermions \cite{Gepner:1987sm} should be useful. 
Another extension would be to the non-compact model, i.e.~the SL(2,$\bbR$) WZNW model. 
In this case, even in the set-up in section \ref{sec:self-dual-case}, Poisson--Lie T-duality may not be self-dual in the standard description. 
This happens when the standard metric and the $B$-field in the dual model become singular in spite that the generalized metric $\cH_{MN}$ is still well-defined. 
Concrete examples of such backgrounds have been studied in \cite{Sakatani:2021eqt} by using certain six-dimensional Drinfel'd doubles. 

Such a class of backgrounds is known to form the non-Riemannian geometry discussed in \cite{Lee:2013hma,Morand:2017fnv}.
It would be interesting to study the quantum aspects of the associated string sigma models.
For the quantization of strings on a flat non-Riemannian geometry, we refer to \cite{Park:2020ixf}.
We also note that, when the condition \eqref{eq:strong-CC} is satisfied as discussed in Appendix \ref{app:lambda} and in section 3 of \cite{Sakatani:2021eqt}, we can find the conserved left and right currents associated with the generalized Killing vector fields $V_{\hat{\gga}}{}^M$ and $K_{\hat{\gga}}{}^M$.
Their associated currents satisfy the same current algebras as those of the SL(2,$\bbR$) WZNW model, and thus would be useful to study the strings on non-Riemannian geometries in the quantum regime.

We may also consider the strings on Poisson--Lie T-fold backgrounds, where the transition functions involve Poisson--Lie T-duality transformations such as the change in the sign of $\bar\varphi$ in our example.
Given the world-sheet description of the duality, it would be possible to study the strings on this class of non-geometric backgrounds by exact conformal field theories.
A simple set-up is to tensor a circle and the group manifold where the latter receives the Poisson--Lie T-duality transformation as a monodromy around the circle.
From the world-sheet point of view, the model which describes the strings there is the orbifold model by the discrete symmetry of the translation along the circle direction accompanied by the duality transformation. 
The world-sheet formulation of such strings, their quantum aspects and possible applications would be explored for example along the line of \cite{Satoh:2015nlc,Satoh:2016izo,Satoh:2022xcs}. 

\vspace{5ex}
\begin{center}
{\large\bf Acknowledgments}
\end{center}

\vspace{-1ex}
We would like to thank Y.~Koga for useful discussions.
This work is supported in part by JSPS KAKENHI Grant Numbers JP22K03631 and JP23K03391.

\vspace{3ex}
\appendix
\renewcommand{\theequation}{\Alph{section}.\arabic{equation}}

\section{Specific case}
\label{app:lambda}
To supplement the discussion in section \ref{sec:CA-WZW}, in this appendix we explain a specific class of $\cE$-models where both the left and the right currents of the WZNW model can be easily constructed from those of the $\cE$-model. 
For an application, see the discussion in the last section of the conclusions.

On this class of the models, we impose a restriction 
\begin{align}
 F_{\ubar{A}\ubar{B}\bar{C}}=0\,,\qquad
 F_{\bar{A}\bar{B}\ubar{C}}=0\,,
\label{eq:strong-CC}
\end{align}
which is called 
the strong conformal condition in \cite{Lacroix:2023ybi}.
At least when $\beta^{ab}=0$ and $\tilde{g}^{ab}$ is non-degenerate, this condition results in the following constraints on the structure constants
\begin{align}
 f_{ab}{}^c = 0\,,\qquad
 f_a{}^{bc} = f_{ade}\,\tilde{g}^{bd}\,\tilde{g}^{ce} \,,
\end{align}
which have been discussed for example in \cite{Demulder:2018lmj} (see Eq.~(5.50)) in the study of the $\lambda$-model. 
If the strong conformal condition is satisfied, the generalized vector fields $K_{\hat{\gga}}{}^M$ defined in Eq.~\eqref{eq:K-def} satisfy
\begin{align}
 [K_{\hat{\gga}},\,K_{\hat{\ggb}}]_{\text{D}}^M = + f_{\hat{\gga}\hat{\ggb}}{}^{\hat{\ggc}}\,K_{\hat{\ggc}}{}^M\,,\qquad
 \hat{\pounds}_{K_{\hat{\gga}}} \cH_{MN}=0\,.
\end{align}
Then the currents $\cJ'_{\hat{\gga}\mp}$ of Eq.~\eqref{eq:cJp} play the role of the right/left currents in the WZNW model, and we do not need to find the matrix $\Lambda_{\hat{\gga}}{}^{{\hat{\ggb}}}$ in Eq.~\eqref{eq:Jtilde}. 
We note that the condition \eqref{eq:strong-CC} is covariant under Poisson--Lie T-duality, and if it is satisfied in the original model, so is in the dual model as well.

For example, in the case of the SU(2) WZNW model, we can choose $\tilde{g}_{ab}=\delta_{ab}$ and
\begin{align}
 f_a{}^{bc} = \epsilon_{ade}\,\delta^{bd}\,\delta^{ce} \,,\qquad
 f_{abc} = \epsilon_{abc}\,.
\end{align}
Then the condition \eqref{eq:strong-CC} is satisfied and we can construct the left and the right currents. 
However, in this example, non-trivial maximally isotropic subalgebras other than the original one generated by $T^a$ are yet to be found, which may not exist in the SU(2) case.
We thus refrain from discussing Poisson--Lie T-duality of this model.

\end{document}